\RequirePackage{rotating}
\documentclass[a4paper,fleqn,usenatbib]{mnras}

\usepackage[T1]{fontenc}
\usepackage{ae,aecompl}
\usepackage{comment}

\usepackage{graphicx}	
\usepackage{amsmath}	
\usepackage{pdflscape}
\usepackage{newtxtext,newtxmath}

\newcommand\nodata{ ~$\cdots$~ }%
\newcommand\gaia{\textit{Gaia\/}~}
\newcommand\hii{H{\sc ii}~}

\title[\gaia distances to Galactic X-ray binaries]{Distances to Galactic X-ray Binaries with \gaia DR2}

\author[R. M. Arnason et al.]{
R. M. Arnason,$^{1}$\thanks{E-mail:rarnaso@uwo.ca}
H. Papei$^{1}$,
P. Barmby$^{1,2}$,
A. Bahramian,$^{3}$
M.D. Gorski$^{1,4}$
\\
$^{1}$Department of Physics \& Astronomy,
$^{2}$Institute for Earth and Space Exploration, University of Western Ontario, 1151 Richmond Street, London, ON N6A 3K7, Canada\\
$^{3}$International Centre for Radio Astronomy Research, Curtin University, GPO Box U1987, Perth, WA 6845, Australia\\
$^{4}$Department of Space, Earth \& Environment, Astronomy and Plasma Physics, Chalmers University of Technology 412 96 Gothenburg, Sweden\\
}

\date{Accepted XXX. Received YYY; in original form ZZZ}

\pubyear{2021}

\begin{document}
\label{firstpage}
\pagerange{\pageref{firstpage}--\pageref{lastpage}}
\maketitle

\begin{abstract}
Precise and accurate measurements of distances to Galactic X-ray binaries (XRBs) reduce uncertainties
in the determination of XRB physical parameters.
We have cross-matched the XRB catalogues of \citet{Liu06a, Liu07a} to the results of \gaia Data Release 2. 
We identify 86 X-ray binaries with a \gaia candidate counterpart, of which 32 are low-mass X-ray binaries (LMXBs) and 54 are high-mass X-ray binaries (HMXBs). 
Distances to \gaia candidate counterparts are, on average, consistent with those measured by \textit{Hipparcos} and radio parallaxes.
When compared to distances measured by \gaia candidate counterparts, distances measured using Type I X-ray bursts are systematically larger, suggesting that these bursts reach only 50\% of the Eddington limit.
However, these results are strongly dependent on the prior assumptions used for estimating distance from the \gaia parallax measurements.
Comparing positions of \gaia candidate counterparts for XRBs in our sample to positions of spiral arms in the Milky Way, we find that HMXBs exhibit  mild preference for being closer to spiral arms; LMXBs exhibit mild preference for being closer to inter-arm regions.
LMXBs do not exhibit any preference for leading or trailing their closest spiral arm. HMXBs exhibit a mild preference for trailing their closest spiral arm.
The lack of a strong correlation between HMXBs and spiral arms may be explained by star formation occurring closer to the midpoint of the arms, or a time delay between star formation and HMXB formation manifesting as a spatial separation between HMXBs and the spiral arm where they formed.
\end{abstract}


\section{Introduction}

\subsection{X-ray Binaries}
\label{xraybinaries}
X-ray binaries (XRBs) are rare systems comprised of a main-sequence star in a close binary orbit with a neutron star (NS) or black hole (BH). 
The accretion of material from the main-sequence companion onto the compact object results in X-ray emission which dominates much of the point source population of the X-ray sky.
Aside from the type of accretor, XRBs are principally categorized based on the mass of the companion.
Binaries where the compact object accretes from the wind of a star $>10$~M$_{\odot}$ are classified as high-mass X-ray binaries (HMXBs), while those that accrete from the Roche lobe overflow of a $<1$ M$_{\odot}$ companion are known as low-mass X-ray binaries \citep[LMXBs;
][]{vanParadijs98a,Casares17a}.
There are a handful of XRBs where the companion is of intermediate mass $1 - 3$~M$_{\odot}$, but they are rare compared to the other two types of system.
It is expected that many primordial intermediate-mass X-ray binaries have evolved to LMXBs in the present day through mass transfer \citep{Podsiadlowski00a}.

XRBs are interesting extraterrestrial laboratories that permit the testing of our understanding of physical processes under extremes of gravity, rotation rate, pressure, temperature, and magnetic field strength. 
In addition, a number of interesting astrophysical phenomena can be studied through XRBs, such as wind physics, neutron star equation of state, and high-energy radiative processes.
Aside from their value to these astrophysical questions, XRBs can also provide independent constraints on their formation environment on larger scales \citep{Lehmer10a,Boroson11a,Zhang12a,Tremmel13a}.
LMXBs can act as independent tracers of stellar mass, since low-mass stars comprise the bulk of the stellar mass in a population \citep{Gilfanov04a}. 
Additionally, LMXBs are preferentially found in areas of high stellar density, such as the globular clusters of the Galaxy and in the direction of the Galactic centre, likely due to their formation by dynamical mechanisms \citep{Clarke75a,Pooley03a,Muno05a,Verbunt06a,Degenaar12a}. 
By contrast, the high-mass companions of HMXBs are short-lived, so they are useful for tracing recent star formation in a long-term galactic evolution context.
Observations of nearby galaxies have suggested that the star formation rate (SFR) of a galaxy scales with both the number of HMXBs and their collective X-ray luminosity, albeit with a moderate dispersion \citep{Grimm03a,Mineo12a}. 
Finally, XRBs are one of the few ways to observe the high mass end of the initial mass function in an evolved population, since isolated neutron stars and black holes are challenging to observe and study \citep{Verbunt87a,Verbunt03a,Dabringhausen12a}.

\subsection{X-ray Binaries and Galactic Structure}
Although field Milky Way XRBs can often be easier to study because of their close proximity (compared to XRBs in globular clusters or other galaxies), investigating the relationship between XRBs and galaxy parameters for the Milky Way is complicated.
Our location within the disc of the Milky Way means that lines of sight where XRBs are expected to be more abundant tend to be heavily extincted.

XRBs tend to have a spatial distribution that is distinct from ordinary stars belonging to the same parent stellar population because the supernova that forms the compact object in an XRB system can impart a velocity kick to the system, often known as a ``natal'' kick.
This velocity kick has two effects: it gives the XRB system a peculiar velocity relative to galactic rotation, and it can substantially displace the system (depending on XRB type) from the star forming region where its progenitor formed \citep{GonzalezHernandez05a,Dhawan07a}.
\cite{Repetto12a} investigated how natal kicks at the birth of black hole LMXBs are necessary to explain their observed distribution in the Milky Way, particularly the presence of  LMXBs at significant (1 kpc) distances above the disc.
They found that these kicks tend to be similar to those found for neutron stars, a property which has been interpreted as a consequence of the asymmetry of the supernova explosion \citep{Janka13a}.

Naively, we expect that if HMXBs are correlated with star formation on a global scale, they should have a spatial correlation with the sites of star formation in the spiral arms. 
The shape and extent of the Milky Way's spiral arms is not easy to resolve compared to external galaxies observed face-on. 
Positions of the spiral arms themselves are typically inferred through the fitting of analytical models to an ensemble of observational tracers, including CO maps, \hii  regions, pulsars, masers, stellar kinematics, and dust emission \citep{Vallee14a}.
To date, investigations of the correlation between HMXBs and the spiral arms have been done using only two proxies of the spiral arms.
\cite{Bodaghee12a} measured spatial cross-correlation between OB associations and HMXBs, finding that they have a characteristic offset of $0.4 \pm 0.2$~kpc, which is attributed to natal kicks received by HMXBs at their formation.
However, by the same models they find no correlation between either OB associations or HMXBs and the spiral arms themselves, which is unexpected given that OB associations are expected to trace out the spiral arms \citep{Brown99a}.
\cite{Coleiro13a} investigated the spatial relation between HMXBs and star forming complexes (SFCs) finding that they are correlated on two characteristic scales: $0.3 \pm 0.05$~kpc and $1.7 \pm 0.3$~kpc, which they interpret as the cluster size and cluster separation, respectively. 
They also derive a mean migration distance for HMXBs of roughly $0.1$~pc and mean migration ages of around $50$~Myr (depending on HMXB type) though they note that sample sizes are small and uncertainties are large.
A large source of that uncertainty lies in the difficulty in determining distances to XRBs within the Milky Way.

\subsection{X-ray Binary Distances}

A principal reason for desiring accurate distances to XRBs in the Milky Way is that many of these XRBs can be studied in detail. 
With the exception of XRBs located in the direction of the Galactic centre, in the Milky Way the population of XRBs can be studied to fainter X-ray luminosities, and identifications of a unique optical counterpart are more straightforward.
Since individual XRBs are most easily studied in the Milky Way, our understanding of individual XRBs in other galaxies and their parameters as an ensemble population are affected by studies of nearby XRBs. 
Measuring the distance to individual XRBs accurately is important because the uncertainty on a number of desired properties in an XRB system can be limited by the uncertainty on distance. 
For example, measurements of distance can affect the inferred size of the accretor (i.e., neutron star radius), inferred mass of either component of the system (either the companion mass or the mass of the accreting neutron star/black hole), inferred mass transfer rate, and other relevant accretion physics due to the inferred luminosity \citep{Galloway03a,Jonker04,Nattila17a,Steiner18a}.

The principal difficulty in measuring distances to XRBs is that they lack a universal property or characteristic that would allow them to be used as a standard candle.
XRBs are also extremely rare compared to ordinary stars, meaning that population-based methods of determining distances to objects, such as main sequence fitting of a star cluster, cannot be used on XRB populations.
Although one can use the main sequence of ordinary stars in a cluster to determine the distance to XRBs in that cluster, the rarity of XRBs means that constructing an ``XRB main sequence" is untenable.
X-ray emission from the accretor which irradiates the companion may modify the expected emission at longer wavelengths, causing an excess in the bluer filters of the visible domain \citep{Phillips99a,MunozDarias05a,Linares18a,Bozzo18a}.
Failing to account for these effects on the expected optical emission of an XRB may lead to incorrect estimates of distance from photometric methods.
These effects are themselves modified by the mass transfer rate, accretion geometry, orbital phase, and accretion state of the system, meaning that they can change with time and may require simultaneous multiwavelength observations for distances to be usefully constrained.

A number of techniques have been used to constrain distance measurements of Milky Way XRBs. 
The most common of these is to measure a photometric distance by assuming that the emission is dominated by the companion at longer wavelengths. 
In general, this method is subject to substantial uncertainties, not only due to the contribution of the accretor, but also due to uncertainties in spectral classification and calibrating the absolute magnitude \citep{Reig15a}. 
A small number of XRBs have had their distances determined via radio parallax or the proper motion of a launched jet \citep{1981ApJ...246L.141H,1999ApJ...512L.121B,2009ApJ...706L.230M}.
This form of measurement provides relatively tight constraints on distance, but is possible only for objects that are sufficiently radio-bright and moderately nearby.

An X-ray specific method of measuring distances is to use the observed flux from Type I X-ray bursts. 
These bursts occur when a sufficient amount of accreted material, mostly hydrogen, accumulates on the surface of a neutron star to trigger a thermonuclear runaway that produces a characteristic burst \citep{Lewin93a}.
The burst is specifically the result of nuclear burning on the neutron star.
A subset of these bursts have steady hydrogen burning followed by ignition of a helium layer beneath the hydrogen layer on the surface. 
The ignition of this helium layer produces a burst that is sufficient to lift the photosphere off the surface of the neutron star.
These bursts are known as photospheric radius expansion (PRE) bursts, and the luminosity of the X-ray burst is expected to be at the Eddington luminosity during the expansion and contraction of the photosphere \citep{Kuulkers03a}.
Since the Eddington limit is fixed for a particular accretor mass (and gas composition/opacity), this means that the mass, radius, and distance of a neutron star can be constrained by comparing the observed flux to the modelled Eddington luminosity for that object.
\citep{Strohmayer06a,Bhattacharyya10a}.
The use of X-ray bursts to infer distance was suggested not long after the detection of such bursts by early X-ray satellites.
This relation has been calibrated using X-ray bursts observed in Galactic globular clusters \citep{vanParadijs78a,vanParadijs81a,Verbunt84a} and applied to several Galactic XRBs that exhibit either PRE or PRE-like bursts \citep{Basinska84a,Galloway03a,Jonker04a}.
Evaluations of this method have shown that uncertainties around the modelling assumptions in this method can result in uncertainties in distance, neutron star mass, and neutron star radius \citep{Galloway08a}. 

With the exception of Type I X-ray bursts, most of the distance-determination techniques require the identification of an optical/infrared counterpart to the X-ray source.
Identification of a counterpart requires high spatial resolution and accurate determination of X-ray position.
Existing catalogues of XRBs include sources which have not been re-detected since their discovery prior to the era of high angular resolution telescopes, and as such have poorly-determined positions that could have many candidate counterparts.
The presence of interstellar extinction along particular lines of sight can interfere with the identification of optical counterparts for many XRB sources. 
Aside from studies of individual objects using telescopes such as the \textit{Hubble Space Telescope}, the principal existing parallax survey of objects in the Milky Way was conducted by the \textit{Hipparcos} satellite \citep{1997AA...323L..49P}.
\textit{Hipparcos} provides parallax for only $\sim10^5$ sources, and has a fairly shallow limiting magnitude of 12.
A handful of nearby XRBs have had their distances determined via \textit{Hipparcos} parallax (see, for example, \citealt{1998AA...330..201C}).
\textit{Hipparcos} data provides reliable measurements of distance within a few hundred parsecs of the Sun, which excludes (based on estimates using the other distance methods described above) the overwhelming majority of XRBs known in the Milky Way. 

\subsection{\gaia DR2 as a Probe of XRB Distances}

The successor to \textit{Hipparcos}, the \gaia satellite, was launched in 2013 and aims to have full five-parameter measurements (position, proper motion, parallaxes) for $\sim1$ billion stars and parallaxes accurate to 10\% for approximately 100 million sources by the end of its five-year mission \citep{gaia2016a}.
To date, there have been two full data releases of \gaia results and an early release of a third version \citep{gaiaDR1,gaiaDR2,gaiaEDR3}. 
\gaia data release 2 (DR2), released in April 2018 and based on the first 22 months of data taken, contains over 1.3 billion sources which have full five-parameter measurements, an improvement of five orders of magnitude of \textit{Hipparcos} for parallax measurements. 
Depending on the required uncertainties, \gaia DR2 contains measurements for objects to a limiting $G$ magnitude of 17--21. 
So far, \gaia DR2 has already provided a wealth of information for studying populations in and nearby the Milky Way that deviate from the expected dynamics of ordinary stars in the Milky Way. 
For example, measurements of candidate hypervelocity stars using \gaia DR2 have shown that many of them are in fact bound to the Milky Way, but at least one object has an origin in the direction of the Magellanic Clouds, suggesting the presence of a supermassive black hole in the Large Magellanic Cloud \citep{Boubert18a,Erkal19}.
\cite{Gandhi19} searched for \gaia DR2 candidate counterparts for Galactic black hole transients, finding that distances from \gaia counterparts generally agreed with prior distance estimates.
Notably, they found that the black hole BW~Cir has a \gaia distance of $\sim 0.6\pm0.2$~kpc, making it the closest dynamically-confirmed transient black hole, although this distance is difficult to reconcile with interpretations of the properties of the donor star.

In this work, we seek to expand the use of \gaia to measure distances to XRBs and assess the accuracy of pre-\gaia distance measurements.
We include not only binaries with black holes/black hole candidates but also neutron star/neutron star candidate binaries and those with no clear identification of accretor type.
Given that XRBs are expected to deviate from the Milky Way's stellar distribution in subtle to dramatic ways, \gaia DR2 offers a unique chance to create a sample of XRBs whose distances are determined by a uniform method, as compared with the heterogeneous mix of methods used for XRB distance determination whose accuracies, systematics, and model dependencies may vary greatly. 
It also offers an opportunity to calibrate alternative methods of measuring distance for use in the general case where parallax measurements are not available. 
\gaia DR2 measurements are subject to several known systematic effects, including centroid wobble caused by unresolved stellar companions \citep{Belokurov2020} and variation in the parallax zero-point with source colour and spatial location \citep{lindegren2018, arenou2018}.
However the widespread use of \gaia data means that these systematics have been investigated and characterized by many different groups \citep[e.g.,][summarize many determinations of the zero-point offset]{chan2020}. 
Uncertainties and systematics are, in general, more poorly understood for the one-off distance measurements available in the literature for many XRBs.

\section{Sample and Methods}

Cross-matching XRBs to \gaia requires input catalogue(s) of known XRBs and XRB candidates. 
To date, the most comprehensive catalogues of XRBs in the Milky Way are the catalogues of high-mass and low-mass XRBs by \citet{Liu06a, Liu07a}.
In general, properties of these XRBs (including positional uncertainties) are compiled using the best/most recent (at the time of catalogue creation) observations of these objects.
These catalogues are assembled from published observations taken with a variety of X-ray telescopes, including \textit{Uhuru}, \textit{Einstein}, \textit{ROSAT}, \textit{RXTE}, \textit{Chandra}, and \textit{XMM-Newton}.
As such, the specific X-ray energies sampled, sensitivities, and coverage of these catalogues are non-uniform. 
Since the most recent updates to these catalogues were in 2006 and 2007, they do not include a number of Galactic XRB candidates discovered since then.
However, an advantage of these catalogues is that many of these objects have been studied in detail, especially those with identified counterparts.
This implies that the expected number of non-XRB contaminants should be low.

\subsection{XRB Sample}
In order to assemble a sample of XRBs for \gaia counterpart matching, we combine the \cite{Liu06a} and \cite{Liu07a} catalogues of Galactic HMXBs and LMXBs. 
Although the most recent revision of these catalogues is now over a decade old, they still represent the most complete sample in the literature.
In total, these catalogues contain 301 XRBs or XRB candidates. 
We have removed two objects from the Liu catalogues: 1H~0556+286, and 1H~1255-567 (Mu-2 Cru), on the basis that they appear to have been misclassified as HMXBs and are in fact ordinary stars \citep{Berghoefer96a,Torrejon01a}. 
The majority of the objects have positional accuracies (equivalent 90 percent confidence) $\sim1$\arcsec or better, typically through identification of an optical counterpart or high-resolution X-ray observation. 
However, a number of the candidate objects in these catalogues have poorly determined positions, especially those that have not been re-observed since the beginning of the \textit{Chandra} era. 
We assume that long-wavelength counterparts identified in the catalogues are true counterparts to the LMXB/HMXB or LMXB/HMXB candidates.
In order to feasibly attempt to identify \gaia counterparts, we select only objects whose positional accuracy is quoted in the catalogues as better than $<10$\arcsec, which provides a sample of 220 XRBs, of which 136 are LMXBs and 84 are HMXBs.

\subsection{Published distance estimates}

Distances to XRBs are estimated using many different methods and a goal of this work is to evaluate the quality of these methods \citep[see also][]{Jonker04,Thevenin2017}.
The \cite{Liu06a} and \cite{Liu07a} catalogues provide distance estimates in the notes to the main catalogue files.
We include the previous distances and original references, as well as an indication of the distance measurement method in Tables~\ref{tab:lmxbtab} and \ref{tab:hmxbtab}, for LMXBs and HMXBs matched to \gaia sources, respectively.
In some cases, only a  distance range is quoted in the Liu et al. catalogues and we give the centre of this range.
In cases where an upper or lower limit was given, we quote that number as the distance.
Fifteen of the XRBs with \gaia candidate counterparts (\autoref{sec:dist_final_sample}) had no previous distance measurement as of the Liu catalogues.
The results of our literature search for these objects are discussed in Appendix~\ref{sect:apdxA}; we found published distances for ten of these fifteen objects. We also updated distances for fifteen additional objects that had more recently-published distances than those given in the Liu catalogues.

The majority of the objects in our sample have distances measured through photometry of the companion, using measured apparent magnitude and extinction with an assumed absolute magnitude based on modelling. 
Many XRBs with a neutron star have had their distance measured using Type I X-ray bursts. 
Aside from these categories, there are also a handful of objects with \textit{Hipparcos} or radio parallaxes, and a variety of other methods for individual objects.
We use the following labels for different distance methods:
\begin{itemize}
\item phot: photometric distance using apparent magnitude, extinction, and assumed absolute magnitude of companion
\item SEDfit: broad-band SED is fit to an assumed model of the companion star/accretion disc with distance as a fitted parameter
\item $A_V$: distance measured using extinction models/Galactic column density
\item jetPM: distance measured using jet proper motion
\item cluster: distance is assumed to be that of an associated cluster/OB association
\item burst: X-ray burst is used as a standard candle to obtain distance
\item VLBAPLX/VLBIPLX: parallax measured using radio interferometery
\item Kin: distance inferred from the kinematics of associated \hii regions
\item HipPLX: distance measured using parallax from the \textit{Hipparcos} satellite
\item unknown: no previous distance measurement
\end{itemize}

In the literature, uncertainties on distances to XRBs are reported in different ways, including making approximations with no quoted uncertainties.
As such, in this work we do not attempt to track the uncertainties associated with previous measurements, except for a handful of cases.
In particular, we expect that distances from a radio parallax (VLBI or VLBA) measurement should be more precise than those from \gaia DR2, and \gaia DR2 distances should agree with parallaxes measured with the \textit{Hipparcos} satellite.
In general we expect that the distance to \gaia candidate counterparts is more reliable and that the \gaia DR2 methodology and systematics are, when taken as a whole, better understood than for the heterogeneous ensemble of other methods.

\subsection{Cross-matching}
\label{sect:xmatch}

We searched for counterparts to our XRB sample by cross-matching with the \gaia DR2 public release. 
Initially, we collected all potential counterparts with a tolerance of $<10$\arcsec and then refined the matches to only include counterparts whose angular separation was less than the quoted positional uncertainty for each individual object.
As per the catalog description, any object that does not have a quoted positional uncertainty is assumed to be accurate to $\sim1$\arcsec  or better \citep{Liu06a,Liu07a}. We have chosen the conservative case of a $\sim1$\arcsec positional uncertainty for these objects.
In the case that an object had asymmetric positional uncertainties in right ascension versus declination, we conservatively chose the maximum of these two.
With this refinement, 99 XRBs from the Liu catalogues have at least one candidate \gaia counterpart.
In total, we find 126 potential counterparts for the Liu XRBs.
Most objects have only one counterpart, while a handful (those with more poorly determined positional accuracy) return two or more potential counterparts. 

We further refined our sample of potential XRB counterparts by considering the probability that each \gaia source is aligned with the position of the XRB by chance alone. 
To estimate probability of our X-ray sources matching a random \gaia source, we picked 5000 random coordinates within 0.1 deg of each X-ray source and crossmatched these random realisations against the Gaia catalog to identify the closest real Gaia source to each random pair of coordinates and measured the angular distance between each random pair of coordinates and the closest real \gaia source to that pair. The distribution of these distances in the vicinity of each X-ray source (for which these random samples were generated and crossmatched against Gaia) is directly proportional to the probability of chance overlap between a source in the Gaia catalog and any random pair of coordinates (informed partially by the density of Gaia catalog in the vicinity of each X-ray source). We approximated the probability of a random match by the fraction of random points which are located within a distance of a \gaia source equal to the separation between the X-ray source and the candidate \gaia counterpart.
After removing the counterparts with a probability of chance overlap greater than 10\%, we obtain 88 \gaia candidate counterparts to the Liu XRB sample, most of which have reported parallaxes.%
\footnote{Except for 2S~0053+604 (see Appendix~\ref{sect:apdxA}), the counterparts without parallaxes are faint ($18.8<G<21.0$) and the lack of parallax measurement is consistent with the distributions given by \cite{gaiaDR2}.
}
A complete list of \cite{Liu06a,Liu07a} catalog sources that were excluded from the final sample and the step at which they were excluded is found in the online supporting information. 
At this level, only two objects have more than one potential \gaia counterpart: AX~J1639.0-4642 and SAX~J1711.6-3808. 
Each of these objects has one potential counterpart with a parallax, and one without.
In the case of AX~J1639.0-4642, the counterpart with parallax is the more probable and we retain that parallax for our analysis. 
The opposite is true for SAX~J1711.6-3808; as for other objects where the \gaia counterpart does not have a parallax, it does not feature in our further analysis.
We searched the literature for more recent distance determinations and any changes to the XRB-type classification for all of the 88 matches.
We found no strong evidence to reclassify individual source types but did find a few additional published distances (see Appendix~\ref{sect:apdxA}).

Before proceeding, we consider the potential biases of our sample compared to the unmatched sample. 
HMXBs have more luminous main-sequence components and unsurprisingly
are more likely to have a counterpart than LMXBs:
there are 187 LMXBs and 114 HMXBs in the Liu catalogues, but we find only 33 and 55 \gaia candidate counterparts to LMXBs and HMXBs, respectively (these numbers each decrease by one after removing the extra candidate counterparts as described above).
Our counterpart matching is also more sensitive to objects that are away from the Galactic centre and away from the Galactic plane - the fraction of objects in the Liu catalogue that have a \gaia candidate counterpart is higher in those directions.

\begin{landscape}
\begin{table}
\scriptsize{
\caption{
Properties of \gaia candidate counterparts to Galactic LMXBs
\label{tab:lmxbtab}}

\begin{tabular}{lccrclcrccll}
Names & RA & DEC & P$_{ \mathrm{interloper}}$ & Gaia DR2 ID & ${\theta}_{ \mathrm{sep}}$ & m$_{\mathrm{G,mean}}$ & GOF & d$_{ \mathrm{Gaia}}$ & d$_{ \mathrm{prev}}$ & d$_{ \mathrm{prev}}$ Type & d$_{ \mathrm{prev}}$ Ref \\
 &  &  &  &  & " & $\mathrm{mag}$ &  & $\mathrm{kpc}$ & $\mathrm{kpc}$ &  &  \\
GRO J0422+32/V518 Per & 04 21 42.790 & +32 54 27.10 & 0.0100 & 172650748928103552 & 0.86 & $ 20.85 \pm 0.05 $ & 4.4 & \nodata & 2.49 & SEDfit & {\cite{2003ApJ...599.1254G}} \\
Swift J061223.0+701243.9/- & 06 12 22.600 & +70 12 43.40 & 0.0001 & 1107229825742589696 & 0.29 & $ 21.00 \pm 0.06 $ & 1.6 & \nodata & \nodata & \nodata & \nodata \\
4U 0614+091/V1055 Ori & 06 17 07.400 & +09 08 13.60 & 0.0051 & 3328832132393159296 & 0.65 & $ 18.56 \pm 0.02 $ & 1.4 & $ 3.3^{+1.3}_{-2.4}$ & 3.0 & burst & {\cite{1992AA...262L..15B}} \\
1A 0620-00/V616 Mon & 06 22 44.503 & -00 20 44.72 & 0.0070 & 3118721026600835328 & 0.73 & $ 17.52 \pm 0.01 $ & 3.0 & $ 1.6^{+0.4}_{-0.7}$ & 1.06 & phot & {\cite{2010ApJ...710.1127C}} \\
4U 0919-54/*X & 09 20 26.950 & -55 12 24.70 & 0.0009 & 5310395631783100800 & 0.1 & $ 20.73 \pm 0.02 $ & 3.0 & \nodata & 5.4 & burst & {\cite{2005AA...441..675I}} \\
GS 1124-684/GU Mus & 11 26 26.700 & -68 40 32.60 & 0.0209 & 5234956524083372544 & 0.64 & $ 19.57 \pm 0.01 $ & 0.4 & $ 2.3^{+1.1}_{-3.1}$ & 5.9 & SEDfit & {\cite{2001PhDT..........G}} \\
1A 1246-588/*X & 12 49 39.364 & -59 05 14.68 & 0.0002 & 6059778089610749440 & 0.04 & $ 20.495 \pm 0.008 $ & 1.2 & $ 2.0^{+1.2}_{-2.4}$ & 5.0 & burst & {\cite{2006AA...446L..17B}} \\
4U 1456-32/V822 Cen & 14 58 22.000 & -31 40 08.00 & 0.0077 & 6205715168442046592 & 0.96 & $ 17.865 \pm 0.005 $ & 0.7 & $ 2.1^{+0.6}_{-1.2}$ & 1.3 & burst & {\cite{1980ApJ...241..779K}} \\
3A 1516-569/BR Cir & 15 20 40.900 & -57 10 01.00 & 0.0763 & 5883218164517055488 & 0.87 & $ 17.92 \pm 0.02 $ & 11.5 & $ 6.2^{+2.0}_{-2.9}$ & 9.2 & burst & {\cite{Jonker04}} \\
1E 1603.6+2600/UW CrB & 16 05 45.820 & +25 51 45.10 & 0.0018 & 1315375795016730880 & 0.75 & $ 19.67 \pm 0.01 $ & 2.4 & $ 2.1^{+0.7}_{-1.2}$ & 6.0 & burst & {\cite{2005MNRAS.356.1133H}} \\
H 1617-155/V818 Sco & 16 19 55.070 & -15 38 24.80 & 0.0001 & 4328198145165324800 & 0.22 & $ 12.48 \pm 0.02 $ & 9.4 & $ 2.13^{+0.21}_{-0.26}$ & 2.8 & VLBAPLX & {\cite{1999ApJ...512L.121B}} \\
4U 1636-536/V801 Ara & 16 40 55.500 & -53 45 05.00 & 0.0819 & 5930753870442684544 & 0.87 & $ 18.27 \pm 0.02 $ & 0.5 & $ 4.4^{+1.6}_{-3.1}$ & 6.0 & burst & {\cite{2008ApJS..179..360G}} \\
GRO J1655-40/V1033 Sco & 16 54 00.137 & -39 50 44.90 & 0.0022 & 5969790961312131456 & 0.15 & $ 16.224 \pm 0.006 $ & 4.0 & $ 3.3^{+0.7}_{-1.1}$ & 3.2 & jetPM & {\cite{1995Natur.375..464H}} \\
2A 1655+353/HZ Her & 16 57 49.830 & +35 20 32.60 & <0.0001 & 1338822021487330304 & 0.26 & $ 13.61 \pm 0.02 $ & 10.4 & $ 5.0^{+0.6}_{-0.7}$ & 6.6 & SEDfit & {\cite{1997MNRAS.288...43R}} \\
MXB 1659-298/V2134 Oph & 17 02 06.500 & -29 56 44.10 & 0.0517 & 6029391608332996224 & 0.48 & $ 19.44 \pm 0.04 $ & 0.1 & \nodata & 10.0 & burst & {\cite{2001ApJ...553L.157M}} \\
4U 1700+24/HD 154791 & 17 06 34.520 & +23 58 18.60 & <0.0001 & 4571810378118789760 & 0.09 & $ 6.743 \pm 0.001 $ & 20.0 & $ 0.536^{+0.009}_{-0.009}$ & 0.42 & phot & {\cite{2002AA...382..104M}} \\
3A 1702-363/V1101 Sco & 17 05 44.500 & -36 25 23.00 & 0.0011 & 5976748056765619328 & 0.11 & $ 17.757 \pm 0.008 $ & -0.0 & $ 4.9^{+1.7}_{-2.9}$ & 9.2 & unclear & {\cite{1995ApJ...447L..33V}} \\
SAX J1711.6-3808/- & 17 11 37.100 & -38 07 05.70 & 0.0299 & 5973177495780065664 & 0.99 & $ 21.05 \pm 0.02 $ & 0.5 & \nodata & \nodata & \nodata & \nodata \\
4U 1724-307/Ter 2 & 17 27 33.300 & -30 48 07.00 & 0.0347 & 4058208396397618688 & 0.34 & $ 18.20 \pm 0.02 $ & 17.2 & $ 6.6^{+2.8}_{-4}$ & 9.5 & cluster & {\cite{Kuulkers03a}} \\
3A 1728-247/V2116 Oph & 17 32 02.160 & -24 44 44.00 & 0.0012 & 4110236324513030656 & 0.15 & $ 15.860 \pm 0.009 $ & 20.0 & $ 7.6^{+2.8}_{-4}$ & 4.5 & phot & {\cite{1997ApJ...489..254C}} \\
4U 1735-444/V926 Sco & 17 38 58.300 & -44 27 00.00 & 0.0742 & 5955379701104735104 & 0.96 & $ 17.77 \pm 0.01 $ & 1.0 & $ 5.6^{+2.1}_{-4}$ & 9.1 & burst & {\cite{1998AA...332..561A}} \\
SLX 1737-282/- & 17 40 43.000 & -28 18 11.90 & 0.0816 & 4060255373456473984 & 0.69 & $ 14.602 \pm 0.003 $ & 47.8 & $ 4.5^{+2.1}_{-4}$ & 6.5 & burst & {\cite{2002AA...389L..43I}} \\
EXO 1747-214/star & 17 50 24.520 & -21 25 19.90 & 0.0067 & 4118590585673834624 & 0.17 & $ 20.24 \pm 0.03 $ & 3.0 & \nodata & 11.0 & burst & {\cite{2005ApJ...635.1233T}} \\
Swift J1753.5-0127/- & 17 53 28.290 & -01 27 06.22 & <0.0001 & 4178766135477201408 & 0.04 & $ 16.698 \pm 0.009 $ & 0.6 & $ 5.6^{+1.8}_{-2.8}$ & 6.0 & A$_{V}$ & {\cite{2007ApJ...659..549C}} \\
4U 1755-33/V4134 Sgr & 17 58 40.000 & -33 48 27.00 & 0.0277 & 4042473487415175168 & 0.35 & $ 19.500 \pm 0.007 $ & 3.7 & \nodata & 6.5 & phot & {\cite{1998ApJ...496L..21W}} \\
2A 1822-371/V691 CrA & 18 25 46.800 & -37 06 19.00 & 0.0186 & 6728016172687965568 & 0.52 & $ 15.53 \pm 0.02 $ & 3.7 & $ 6.1^{+1.6}_{-2.7}$ & 2.5 & SEDfit & {\cite{1982ApJ...262..253M}} \\
HETE J1900.1-2455/star & 19 00 08.650 & -24 55 13.70 & 0.0007 & 4074363039644919936 & 0.17 & $ 18.10 \pm 0.01 $ & -0.1 & $ 3.5^{+1.7}_{-3.5}$ & 5.0 & burst & {\cite{2005ATel..534....1K}} \\
4U 1908+005/V1333 Aql & 19 11 16.000 & +00 35 06.00 & 0.0638 & 4264296556603631872 & 0.87 & $ 18.901 \pm 0.004 $ & 2.4 & $ 3.0^{+1.3}_{-2.6}$ & 5.2 & burst & {\cite{Jonker04}} \\
4U 1916-05/V1405 Aql & 19 18 47.870 & -05 14 17.09 & 0.0066 & 4211396994895217152 & 0.37 & $ 20.92 \pm 0.03 $ & 0.6 & \nodata & 8.9 & burst & {\cite{2008ApJS..179..360G}} \\
3A 1954+319/star & 19 55 42.330 & +32 05 49.10 & 0.0014 & 2034031438383765760 & 0.12 & $ 8.370 \pm 0.002 $ & 23.6 & $ 3.3^{+0.6}_{-1.0}$ & 1.7 & phot & {\cite{2006AA...453..295M}} \\
GS 2023+338/V404 Cyg & 20 24 03.830 & +33 52 02.200 & 0.0058 & 2056188620566335360 & 0.24 & $ 17.19 \pm 0.01 $ & 4.0 & $ 2.1^{+0.4}_{-0.6}$ & 2.39 & VLBIPLX & {\cite{2009ApJ...706L.230M}} \\
4U 2129+47/V1727 Cyg & 21 31 26.200 & +47 17 24.00 & 0.0106 & 1978241050130301312 & 0.53 & $ 17.600 \pm 0.001 $ & 1.9 & $ 1.75^{+0.26}_{-0.4}$ & 6.3 & phot & {\cite{1990AJ.....99..678C}} \\
\end{tabular}

\bigskip
First name in each column indicates the first name in the catalogue, while the second name indicates the name of the optical counterpart (if any).
Optical counterparts that have numbers/letters following a * refers to the corresponding object on the finding chart as described in the Liu catalogues.
Optical counterparts with the name "star" do not have a labelled object on their corresponding finding chart. 
 ${\theta}_{ \mathrm{sep}}$ indicates the separation between the candidate \textit{Gaia} counterpart and the quoted position of the XRB in \cite{Liu06a,Liu07a}.
GOF is the \gaia DR2 goodness-of-fit statistic {\tt astrometric\_gof\_al}.
} 

\end{table}

\end{landscape}

\begin{landscape}
\begin{table}
\tiny{
\caption{
Properties of \gaia candidate counterparts to Galactic HMXBs
\label{tab:hmxbtab}
} 
\begin{tabular}{lccrrlcrccll}
Names & RA & DEC & P$_{ \mathrm{interloper}}$ & Gaia DR2 ID & ${\theta}_{ \mathrm{sep}}$ & m$_{\mathrm{G,mean}}$ & GOF & d$_{ \mathrm{Gaia}}$ & d$_{ \mathrm{prev}}$ & d$_{ \mathrm{prev}}$ Type & d$_{ \mathrm{prev}}$ Ref \\
 &  &  &  &  & " & $\mathrm{mag}$ &  & $\mathrm{kpc}$ & $\mathrm{kpc}$ &  &  \\

2S 0053+604/{gamma} Ca & 00 56 42.50 & +60 43 00.0 & 0.0041 & 426558460877467776 & 0.66 & $ 1.82 \pm 0.01 $ & 190.5 & \nodata & 0.19 & HipPLX & {\cite{1997AA...323L..49P}} \\
2S 0114+650/V662 Cas & 01 18 02.70 & +65 17 30.0 & 0.0005 & 524924310153249920 & 0.17 & $ 10.520 \pm 0.001 $ & 5.5 & $ 6.6^{+1.1}_{-1.6}$ & 7.2 & phot & {\cite{1996AA...311..879R}} \\
RX J0146.9+6121/LS I +61 235 & 01 47 00.20 & +61 21 23.7 & 0.0001 & 511220031584305536 & 0.1 & $ 11.210 \pm 0.002 $ & 19.3 & $ 2.50^{+0.18}_{-0.21}$ & 2.3 & phot & {\cite{1993MNRAS.261..599C}} \\
IGR J01583+6713/- & 01 58 18.44 & +67 13 23.5 & 0.0009 & 518990967445248256 & 0.3 & $ 13.7000 \pm 0.0007 $ & -1.1 & $ 7.4^{+0.8}_{-1.1}$ & 6.4 & phot & {\cite{2006AA...455...11M}} \\
1E 0236.6+6100/LS I +61 303 & 02 40 31.70 & +61 13 46.0 & 0.0031 & 465645515129855872 & 0.48 & $ 10.390 \pm 0.001 $ & 3.3 & $ 2.45^{+0.21}_{-0.26}$ & 2.4 & Kin & \nodata \\
V 0332+53/BQ Cam & 03 34 59.90 & +53 10 24.0 & 0.0025 & 444752973131169664 & 0.71 & $ 14.220 \pm 0.002 $ & 8.6 & $ 5.1^{+0.8}_{-1.0}$ & 7.0 & phot & {\cite{1999MNRAS.307..695N}} \\
RX J0440.9+4431/LS V +44 17 & 04 40 59.30 & +44 31 49.0 & 0.0014 & 252878401557369088 & 0.41 & $ 10.430 \pm 0.001 $ & -0.0 & $ 3.2^{+0.5}_{-0.6}$ & 3.2 & phot & {\cite{1997AA...323..853M}} \\
EXO 051910+3737.7/V420 Aur & 05 22 35.20 & +37 40 34.0 & 0.0025 & 184497471323752064 & 0.52 & $ 7.220 \pm 0.001 $ & 11.8 & $ 1.29^{+0.09}_{-0.10}$ & 1.7 & phot & {\cite{1990AA...231..354P}} \\
1A 0535+262/V725 Tau & 05 38 54.60 & +26 18 57.0 & 0.0011 & 3441207615229815040 & 0.37 & $ 8.680 \pm 0.007 $ & 5.6 & $ 2.13^{+0.21}_{-0.26}$ & 2.45 & phot & {\cite{1998MNRAS.297L...5S}},{\cite{2000AstL...26....9L}} \\
IGR J06074+2205/- & 06 07 26.60 & +22 05 48.3 & 0.0039 & 3423526544838563328 & 0.57 & $ 12.180 \pm 0.001 $ & 3.1 & $ 5.5^{+1.0}_{-1.5}$ & 4.5 & phot & {\cite{2010AA...522A.107R}} \\
SAX J0635.2+0533/- & 06 35 18.29 & +05 33 06.3 & <0.0001 & 3131755947406031104 & 0.15 & $ 12.510 \pm 0.006 $ & 0.7 & $ 5.7^{+1.3}_{-2.0}$ & 3.75 & phot & {\cite{1999ApJ...523..197K}} \\
XTE J0658-073/[M81] I-33 & 06 58 17.30 & -07 12 35.3 & 0.0004 & 3052677318793446016 & 0.2 & $ 12.030 \pm 0.003 $ & 4.1 & $ 5.1^{+0.9}_{-1.4}$ & 3.9 & phot & {\cite{2006AA...451..267M}} \\
3A 0726-260/V441 Pup & 07 28 53.60 & -26 06 29.0 & 0.0019 & 5613494119544761088 & 0.3 & $ 11.620 \pm 0.003 $ & 4.5 & $ 9.5^{+2.1}_{-3.1}$ & 5.35 & phot & {\cite{1984AA...131..385C}},{\cite{1996AA...315..160N}} \\
1H 0739-529/HD 63666 & 07 47 23.60 & -53 19 57.0 & 0.0005 & 5489434710755238400 & 0.22 & $ 7.5200 \pm 0.0004 $ & 21.0 & $ 0.643^{+0.017}_{-0.018}$ & 0.52 & HipPLX & {\cite{1998AA...330..201C}} \\
4U 0900-40/HD 77581 & 09 02 06.90 & -40 33 17.0 & 0.0044 & 5620657678322625920 & 0.46 & $ 6.720 \pm 0.002 $ & 9.6 & $ 2.42^{+0.16}_{-0.19}$ & 1.9 & phot & {\cite{1985ApJ...288..284S}} \\
GRO J1008-57/star & 10 09 46.90 & -58 17 35.5 & 0.0052 & 5258414192353423360 & 0.52 & $ 13.900 \pm 0.001 $ & 10.8 & $ 3.6^{+0.4}_{-0.5}$ & 5.0 & phot & {\cite{1994MNRAS.270L..57C}} \\
1A 1118-615/Hen 3-640 & 11 20 57.20 & -61 55 00.0 & 0.0055 & 5336957010898124160 & 0.25 & $ 11.6000 \pm 0.0005 $ & 4.7 & $ 2.93^{+0.22}_{-0.26}$ & 5.0 & phot & {\cite{1981AA....99..274J}} \\
4U 1119-603/V779 Cen & 11 21 15.10 & -60 37 25.5 & 0.0021 & 5337498593446516480 & 0.14 & $ 12.890 \pm 0.003 $ & 8.2 & $ 6.4^{+1.0}_{-1.4}$ & 9.0 & phot & {\cite{1974ApJ...192L.135K}},{\cite{1979ApJ...229.1079H}} \\
IGR J11215-5952/HD 306414 & 11 21 46.81 & -59 51 47.9 & 0.0017 & 5339047221168787712 & 0.12 & $ 9.760 \pm 0.001 $ & 6.3 & $ 6.5^{+1.1}_{-1.5}$ & 6.2 & phot & {\cite{2006AA...449.1139M}} \\
IGR J11435-6109/- & 11 44 10.70 & -61 07 02.0 & 0.0363 & 5335022901224296064 & 0.64 & $ 13.000 \pm 0.002 $ & 52.8 & $ 3.9^{+1.1}_{-1.8}$ & 8.6 & phot & {\cite{2009AA...495..121M}} \\
2S 1145-619/V801 Cen & 11 48 00.00 & -62 12 25.0 & 0.0023 & 5334823859608495104 & 0.18 & $ 8.630 \pm 0.002 $ & 7.1 & $ 2.23^{+0.16}_{-0.19}$ & 3.1 & phot & {\cite{1997MNRAS.288..988S}} \\
1E 1145.1-6141/V830 Cen & 11 47 28.60 & -61 57 14.0 & 0.0594 & 5334851450481641088 & 0.64 & $ 12.280 \pm 0.001 $ & 6.6 & $ 9.1^{+1.6}_{-2.2}$ & 8.0 & phot & {\cite{1982AA...114L...7I}} \\
4U 1223-624/BP Cru ? & 12 26 37.60 & -62 46 13.0 & 0.0075 & 6054569565614460800 & 0.37 & $ 9.760 \pm 0.001 $ & 12.6 & $ 3.5^{+0.4}_{-0.5}$ & 4.1 & phot & {\cite{2002AA...391..219L}} \\
1H 1249-637/HD 110432 & 12 42 50.30 & -63 03 31.0 & 0.0037 & 6055103928246312960 & 0.24 & $ 5.120 \pm 0.002 $ & 76.4 & $ 0.416^{+0.021}_{-0.023}$ & 0.392 & A$_{V}$ & {\cite{2009AA...507..833M}} \\
1H 1253-761/HD 109857 & 12 39 14.60 & -75 22 14.0 & 0.0003 & 5837600152935767680 & 0.16 & $ 6.5200 \pm 0.0004 $ & 22.3 & $ 0.2117^{+0.0014}_{-0.0014}$ & 0.24 & HipPLX & {\cite{1998AA...330..201C}} \\
4U 1258-61/V850 Cen & 13 01 17.10 & -61 36 07.0 & 0.0005 & 5863533199843070208 & 0.36 & $ 12.650 \pm 0.003 $ & 16.0 & $ 2.01^{+0.13}_{-0.15}$ & 2.4 & phot & {\cite{1980MNRAS.190..537P}} \\
2RXP J130159.6-635806/- & 13 01 58.70 & -63 58 09.0 & 0.0029 & 5862285700835092352 & 0.22 & $ 17.340 \pm 0.001 $ & 6.7 & $ 5.5^{+1.7}_{-2.8}$ & 5.5 & phot & {\cite{2005MNRAS.364..455C}} \\
2S 1417-624/*7 & 14 21 12.90 & -62 41 54.0 & 0.0995 & 5854175187681966464 & 0.81 & $ 20.490 \pm 0.006 $ & 0.6 & $ 3.8^{+1.7}_{-2.7}$ & 6.2 & phot & {\cite{1984ApJ...276..621G}} \\
4U 1538-52/QV Nor & 15 42 23.30 & -52 23 10.0 & 0.0267 & 5886085557746480000 & 0.72 & $ 13.190 \pm 0.001 $ & 13.6 & $ 6.6^{+1.4}_{-2.1}$ & 4.5 & phot & {\cite{2004ApJ...610..956C}},{\cite{1992MNRAS.256..631R}} \\
1H 1555-552/HD 141926 & 15 54 21.80 & -55 19 45.0 & 0.0610 & 5884544931471259136 & 0.74 & $ 8.680 \pm 0.001 $ & 4.3 & $ 1.35^{+0.08}_{-0.09}$ & 0.96 & \nodata & {\cite{1992ApJS...81..795G}} \\
IGR J16318-4848/*1 & 16 31 48.31 & -48 49 00.7 & 0.0002 & 5940777877435137024 & 0.04 & $ 17.170 \pm 0.002 $ & 32.4 & $ 5.2^{+1.8}_{-2.7}$ & 0.9 & SEDfit & {\cite{2004ApJ...616..469F}} \\
AX J1639.0-4642/- & 16 39 05.40 & -46 42 14.0 & 0.0149 & 5942638074996489088 & 0.4 & $ 19.800 \pm 0.006 $ & 7.3 & $ 4.0^{+1.7}_{-2.6}$ & \nodata & \nodata & \nodata \\
IGR J16465-4507/- & 16 46 35.26 & -45 07 04.5 & 0.0008 & 5943246345430928512 & 0.11 & $ 13.510 \pm 0.001 $ & 9.0 & $ 2.70^{+0.35}_{-0.5}$ & 12.5 & phot & {\cite{2004ATel..338....1S}} \\
IGR J16479-4514/- & 16 48 07.00 & -45 12 05.8 & 0.0724 & 5940244030149933696 & 1.29 & $ 19.580 \pm 0.005 $ & 3.5 & $ 2.8^{+1.4}_{-2.5}$ & 4.45 & phot & {\cite{2015ApJ...808..140C}} \\
4U 1700-37/HD 153919 & 17 03 56.80 & -37 50 39.0 & 0.0154 & 5976382915813535232 & 0.33 & $ 6.400 \pm 0.001 $ & 12.9 & $ 1.75^{+0.19}_{-0.23}$ & 2.12 & A$_{V}$ & {\cite{2009AA...507..833M}} \\
XTE J1739-302/- & 17 39 11.58 & -30 20 37.6 & 0.0039 & 4056922105185686784 & 0.4 & $ 12.670 \pm 0.001 $ & 34.1 & $ 5.3^{+2.1}_{-4}$ & 2.3 & phot & {\cite{2006ApJ...638..982N}} \\
IGR J17544-2619/*C1 & 17 54 25.28 & -26 19 52.6 & 0.0006 & 4063908810076415872 & 0.1 & $ 11.670 \pm 0.001 $ & 9.5 & $ 2.66^{+0.33}_{-0.4}$ & 3.2 & phot & {\cite{2006AA...455..653P}} \\
SAX J1819.3-2525/V4641 Sgr & 18 19 21.48 & -25 25 36.0 & 0.0168 & 4053096217067937664 & 0.28 & $ 18.82 \pm 0.02 $ & 3.2 & \nodata & 6.2 & SEDfit & {\cite{2014ApJ...784....2M}} \\
RX J1826.2-1450/LS 5039 & 18 26 15.06 & -14 50 54.3 & 0.0004 & 4104196427943626624 & 0.08 & $ 10.8000 \pm 0.0004 $ & -2.6 & $ 1.96^{+0.19}_{-0.23}$ & 2.5 & phot & {\cite{2005MNRAS.364..899C}} \\
AX J1841.0-0536/- & 18 41 00.43 & -05 35 46.5 & 0.0017 & 4256500538116700160 & 0.09 & $ 12.940 \pm 0.003 $ & 34.3 & $ 7.6^{+2.2}_{-3.1}$ & 10.0 & A$_{V}$ & {\cite{2001PASJ...53.1179B}} \\
XTE J1901+014/star & 19 01 39.90 & +01 26 39.2 & 0.0045 & 4268294763113217152 & 0.6 & $ 19.450 \pm 0.007 $ & 0.2 & $ 2.2^{+1.1}_{-2.2}$ & \nodata & \nodata & \nodata \\
XTE J1906+09/star & 19 04 47.48 & +09 02 41.8 & 0.0014 & 4310649149314811776 & 0.23 & $ 19.73 \pm 0.01 $ & 6.8 & $ 2.8^{+1.4}_{-2.3}$ & 10.0 & A$_{V}$ & {\cite{1998ApJ...502L.129M}} \\
3A 1909+048/SS 433 & 19 11 49.60 & +04 58 58.0 & 0.0102 & 4293406612283985024 & 0.56 & $ 12.63 \pm 0.02 $ & 18.1 & $ 3.8^{+0.8}_{-1.1}$ & 5.5 & jetPM & {\cite{1981ApJ...246L.141H}} \\
4U 1909+07/*A & 19 10 48.20 & +07 35 52.3 & 0.0028 & 4306419980916246656 & 0.62 & $ 20.170 \pm 0.008 $ & 10.6 & $ 2.6^{+1.3}_{-2.3}$ & 7.0 & A$_{V}$ & {\cite{2000ApJ...532.1119W}} \\
IGR J19140+0951/- & 19 14 04.20 & +09 52 58.3 & 0.0056 & 4309253392325650176 & 0.41 & $ 18.200 \pm 0.006 $ & 13.7 & $ 2.8^{+1.3}_{-2.3}$ & 3.6 & phot & {\cite{2010AA...510A..61T}} \\
1H 1936+541/DM +53 2262 & 19 32 52.30 & +53 52 45.0 & 0.0009 & 2136886799749672320 & 0.48 & $ 10.370 \pm 0.002 $ & 23.2 & $ 3.3^{+0.4}_{-0.5}$ & \nodata & \nodata & \nodata \\
XTE J1946+274/*A & 19 45 39.30 & +27 21 55.4 & 0.0422 & 2028089540103670144 & 0.76 & $ 15.7100 \pm 0.0007 $ & 6.8 & $ 12.6^{+2.9}_{-4}$ & 9.5 & SEDfit & {\cite{2003ApJ...584..996W}} \\
KS 1947+300/*3 & 19 49 30.50 & +30 12 24.0 & 0.0359 & 2031938140034489344 & 0.57 & $ 20.48 \pm 0.01 $ & -0.0 & $ 3.1^{+2.0}_{-3.5}$ & 9.5 & pulsar & {\cite{2005AstL...31...88T}} \\
4U 1956+35/HD 226868 & 19 58 21.70 & +35 12 06.0 & 0.0051 & 2059383668236814720 & 0.37 & $ 8.5200 \pm 0.0008 $ & 3.6 & $ 2.23^{+0.15}_{-0.18}$ & 1.86 & VLBAPLX & {\cite{2011ApJ...742...83R}} \\
EXO 2030+375/*2 & 20 32 15.20 & +37 38 15.0 & 0.0116 & 2063791369815322752 & 0.9 & $ 16.910 \pm 0.003 $ & 13.5 & $ 3.6^{+0.9}_{-1.3}$ & 7.1 & A$_{V}$ & {\cite{2002ApJ...570..287W}} \\
RX J2030.5+4751/SAO 49725 & 20 30 30.80 & +47 51 51.0 & 0.0125 & 2083644392294059520 & 0.54 & $ 9.0300 \pm 0.0006 $ & 9.1 & $ 2.49^{+0.16}_{-0.19}$ & 2.2 & phot & {\cite{1997AA...323..853M}} \\
GRO J2058+42/star & 20 58 47.50 & +41 46 37.0 & 0.0048 & 2065653598916388352 & 0.45 & $ 14.190 \pm 0.005 $ & -0.3 & $ 8.0^{+0.9}_{-1.2}$ & 9.0 & \nodata & {\cite{2005AA...440..637R}} \\
1H 2202+501/BD +49 3718 & 22 01 38.20 & +50 10 05.0 & 0.0029 & 1979911002134040960 & 0.37 & $ 9.3000 \pm 0.0004 $ & 15.4 & $ 1.16^{+0.05}_{-0.05}$ & 0.7 & HipPLX & {\cite{1998AA...330..201C}} \\
4U 2206+543/BD +53 2790 & 22 07 56.20 & +54 31 06.0 & 0.0086 & 2005653524280214400 & 0.52 & $ 9.7400 \pm 0.0007 $ & 8.5 & $ 3.34^{+0.32}_{-0.4}$ & 2.6 & phot & {\cite{2006AA...446.1095B}} \\

\end{tabular}

\bigskip
First name in each column indicates the first name in the catalogue, while the second name indicates the name of the optical counterpart (if any).
Optical counterparts that have numbers/letters following a * refers to the corresponding object on the finding chart as described in the Liu catalogues.
Optical counterparts with the name ``star'' do not have a labelled object on their corresponding finding chart. 
${\theta}_{ \mathrm{sep}}$ indicates the separation between the candidate \textit{Gaia} counterpart and the quoted position of the XRB in \cite{Liu06a,Liu07a}.
GOF is the \gaia DR2 goodness-of-fit statistic {\tt astrometric\_gof\_al}.

} 
\end{table}

\end{landscape}

\subsection{Distances and Final Sample}
\label{sec:dist_final_sample}
To obtain the distance for each counterpart, we match the \gaia source ID to the catalogue of \cite{BailerJones18a}, which uses a Bayesian method to infer distances.
In this work, we quote distance uncertainties as the 1$\sigma$ bounds on the posterior probability density function for distance.
In general, this function is asymmetric about the peak value, so we have asymmetric error bars. 
The prior of this Bayesian method models the Galactic stellar density as an exponential disc, so 
the particular distance prior assumed for each object depends on that object's position in Galactic coordinates.
For ordinary stars, information such as line-of-sight extinction, measured $T_{\rm eff}$, and magnitude/colours in the \gaia filters can provide additional distance constraints. 
However, for XRBs we prefer the position-plus-parallax-only method used by \cite{BailerJones18a},  since modelling the expected value of the additional other parameters in an XRB system is more complex than for an individual star or ordinary binary. 

Since LMXBs do not follow the same spatial distribution as the stellar distribution assumed by \cite{BailerJones18a}, we must cautiously interpret the distances to LMXBs \citep{Grimm02a}(see \autoref{sec:dist_comp}).
For example, an exponential disc model would prefer smaller distances for objects along lines of sight that are out of the plane of the Milky Way.
However, this may not be optimal for LMXBs, given that they can be displaced from the stellar distribution by supernova kicks. 
Of the matched XRBs, 76 of the Liu catalogue counterparts have a parallax: 24 of these counterparts are associated with LMXBs, while 52 are associated with HMXBs.
As the GOF values in Tables~\ref{tab:lmxbtab} and \ref{tab:hmxbtab} show, the astrometric goodness-of-fit is poor for many of these sources. 
The \gaia DR2 documentation suggests that $|{\rm GOF}|>3$ indicates possible problems with the fit: 12 LMXBs and 44 HMXBs have values in this range. A large subset of sources in our sample (30) appear to show a large excess noise in their astrometric fit (larger than 5-$\sigma$). A reason for the high GOF and large excess noise in these systems could be the orbital motion.

\begin{figure}
\includegraphics[width=\columnwidth]{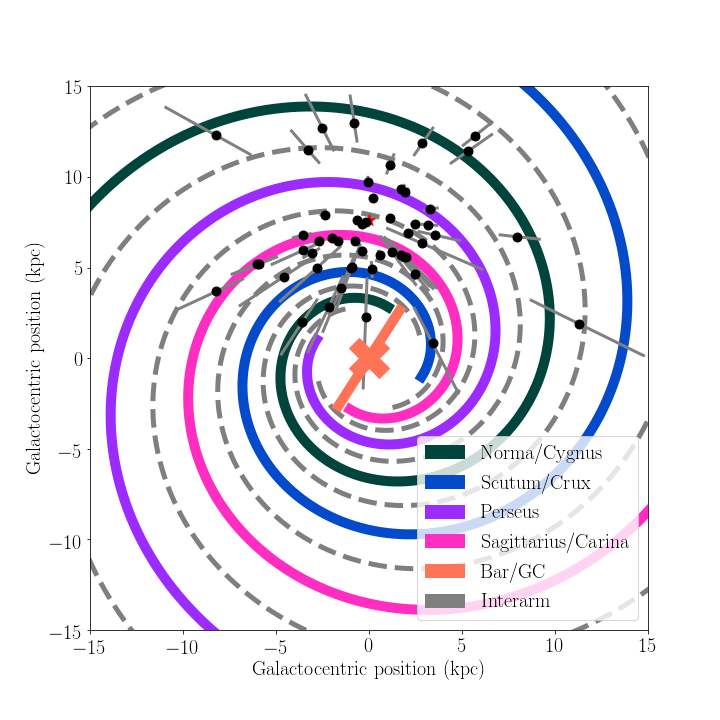}
\caption[Face-on distribution of \gaia counterparts for Liu HMXBs]{
Face-on distribution of \gaia counterparts for Liu HMXBs. 
The spiral arms are modelled using the symmetric spiral arm model of \protect\cite{Vallee08a}.
Interarm regions are modelled as the symmetric arm model phase shifted by 45 degrees.
Error bars for distance/parallax represent the 1$\sigma$ uncertainties. The sun is located at the red star in the middle upper portion of the Figure.}
\label{fig:hmxb_face}
\end{figure}

\begin{figure}
\includegraphics[width=\columnwidth]{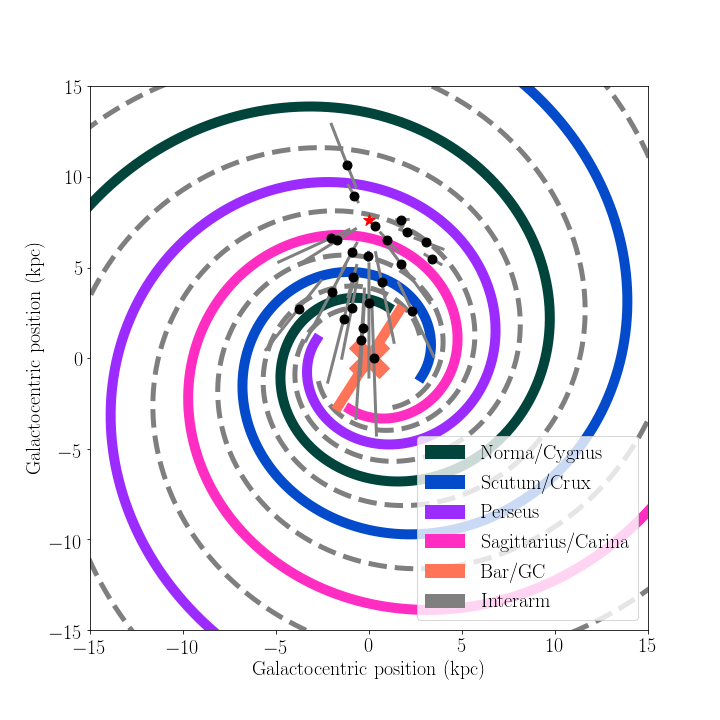}
\caption{Face-on distribution of \gaia counterparts for Liu LMXBs. 
The spiral arms are modelled using the symmetric spiral arm model of \protect\cite{Vallee08a}.
Interarm regions are modelled as the symmetric arm model phase shifted by 45 degrees.
Error bars for distance/parallax represent the 1$\sigma$ uncertainties. The sun is located at the red star in the middle upper portion of the Figure.
}
\label{fig:lmxb_face}
\end{figure}

Several of the matched objects have a negative measured parallax. 
In this case, the distance we obtain is dominated by the assumptions of the prior (see discussion in \citealp{Luri18a} and \citealp{Hogg18a}).
We plot the positions of the \gaia candidate counterparts to the XRBs for a face-on projection of the Milky Way in Figures~\ref{fig:hmxb_face} and \ref{fig:lmxb_face}.

\section{Results}

Several expected results are evident in the Galactic distributions of the XRB sample.
First, as shown in \autoref{fig:hmxb_face}, HMXBs appear to trace out the nearby (i.e., within 5--8 kpc) arms of the Galaxy. 
Since HMXB luminosity is correlated with star formation rate in star-forming galaxies \citep{Grimm03a,Mineo12a}, and spiral arms are the primary sites of star formation, it is reasonable to infer that they should be spatially close to spiral arms.
\autoref{fig:lmxb_face} shows that the LMXBs are preferentially found in the direction toward the Galactic centre.
A Rayleigh test rejects the null hypothesis that the Galactic longitudes of the LMXBs are uniformly distributed at $p=1.5\times10^{-5}$ (and a Kuiper two-sample test rejects the hypothesis that the Galactic longitude distributions of the LMXBs and HMXBs are drawn from the same distribution at $p=0.02$.)
The concentration of LMXBs toward the Galactic centre is also expected since LMXBs have been shown to trace stellar mass in galaxies \citep{Gilfanov04a} and are preferentially formed in dense areas with high stellar encounter rate, such as the Galactic Bulge \citep{Muno05a, Degenaar12a}.

\subsection{Distance Measurement Comparison} 
\label{sec:dist_comp}

\begin{figure*}

\includegraphics[width=\textwidth]{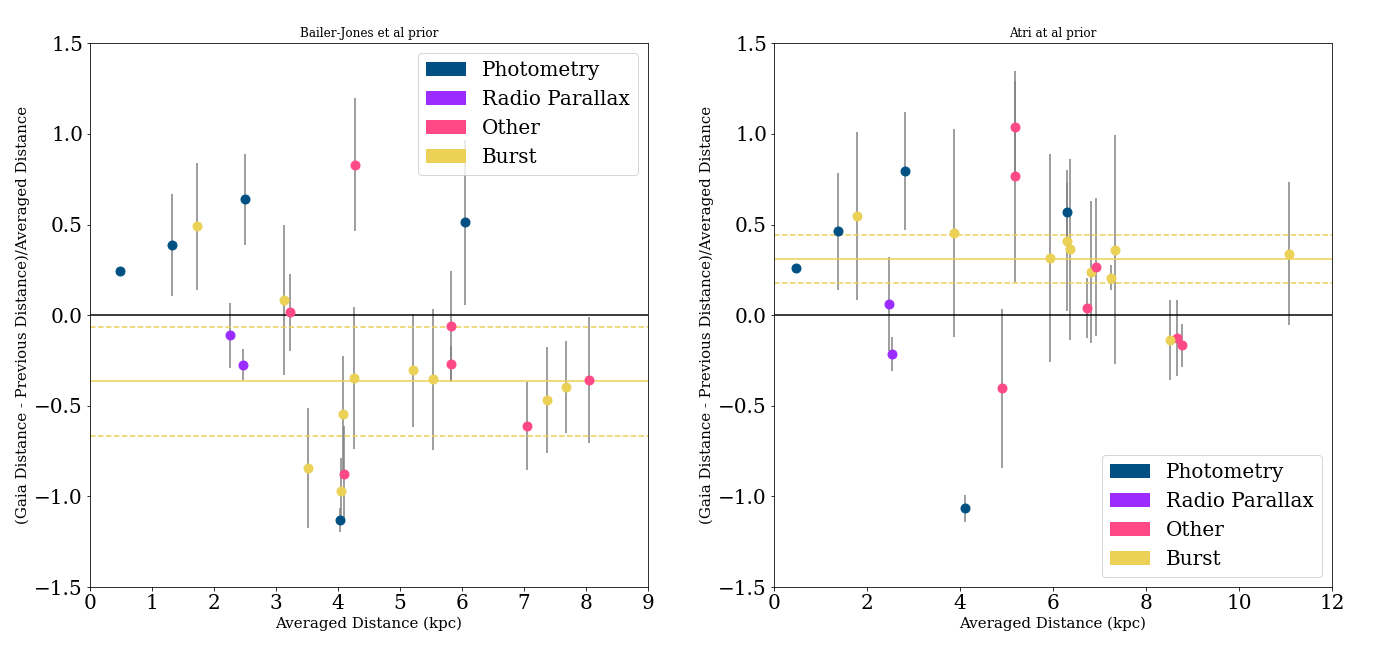}

\caption{
    Difference between literature distance measurements for Liu catalogue LMXBs and the distances obtained in this work, versus the average of these distances. Error bars are lower limits because the uncertainties in the previous distance measurements were omitted. The horizontal yellow lines show the mean and 95\% confidence interval of the sources with previously measured distances using Type I X-ray burst. 
    Left panel: \gaia DR2 distances calculated using the prior of \citet{BailerJones18a}.  
    Right panel: \gaia DR2 distances calculated using the prior of \citet{Atri19}.
    Type I X-ray burst distances show a systematic overestimate with respect to the Gaia distances using the Bailer-Jones prior, and a systematic underestimate with respect to the Gaia distances that use the Atri et al. priors. 
    The comparison using the prior of \citet{BailerJones18a} implies that X-ray bursts (assuming they are PRE bursts) only reach 0.5~L$_{\rm Eddington}$. 
    }
\label{fig:lmxb_dist}

\end{figure*}

\begin{figure}
\includegraphics[width=\columnwidth]{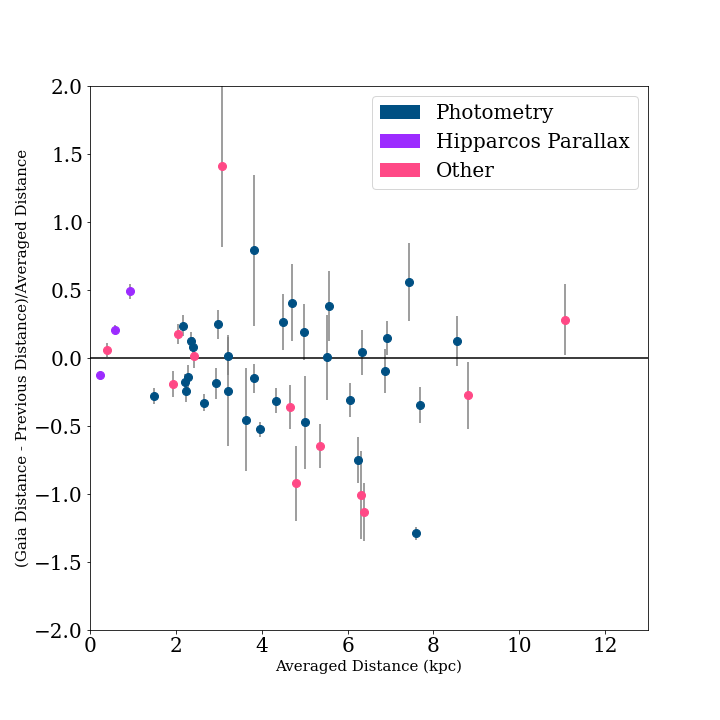}
\caption{
   Difference between previous distance measurements for Liu catalogue HMXBs and the distances obtained in this work, versus the average of these distances.
   Previous distances are obtained from the literature reference given in  \autoref{tab:hmxbtab}, while the distance in this work is the distance to the \gaia candidate counterpart for each HMXB. 
    }
\label{fig:hmxb_dist}
\end{figure}

We find \gaia parallax measurements for less than one third of the combined LMXB/HMXB catalogue; in general, parallax measurements will not be available for Galactic XRBs.
Hence it is useful to use objects with parallax measurement as a diagnostic for other distance methods.
We show a comparison of previous distance measurements with those derived from the \gaia candidate counterparts using Bland-Altman plots in Figures~\ref{fig:lmxb_dist} and \ref{fig:hmxb_dist}. These two figures show the ratio of the difference between previous measured distances and \gaia DR2 distances (this work) to the average of the measurements versus the average of the measurements.
To estimate the error bars in these plots, we only used the uncertainties for our \gaia counterparts, but omit the uncertainties on previous measurements given the difficulties in comparing methods, instruments, and the fact that many distances are assumed rather than measured directly. 
A useful component of this work is to tabulate distance methods for XRBs with \gaia candidate counterparts, since in the Liu catalogue, distances  are reported without specifying the methodology or whether distances are measured or assumed. 

Most previous distance-measurement methods produce distances consistent with \gaia, within their uncertainties where available.
\gaia DR2 measurements agree well with objects whose parallax has been previously measured either by \textit{Hipparcos} or radio interferometery (VLBI/VLBA).
Radio interferometery parallaxes are expected to be significantly more accurate than \gaia; 
therefore, comparing with radio parallaxes verifies the assumptions of the \gaia prior, at least for the distance ranges and directions where objects with radio parallax are available.
In addition, the mean and median differences between distances previously measured photometrically and \gaia DR2 distances are consistent with zero. 

LMXB distances measured using Type I X-ray bursts do show evidence of a trend with a plausible physical interpretation.
As shown in the left panel of \autoref{fig:lmxb_dist}, distances measured using Type I X-ray bursts are systematically larger than those measured via \gaia candidate counterparts.
This figure shows the ratio of difference between previously measured distance and \gaia distance to the average of 
previously measured distance and \gaia distance.
The mean of this quantity over all the methods excluding the Type I X-ray bursts is $-0.07 \pm 0.32$,  consistent with the null hypothesis of no difference (zero mean) with a $p$-value of $0.17$.  However, the mean for the  Type I X-ray bursts is $-0.36 \pm 0.29$ with a $p$-value of $0.01$, which means we can reject the null hypothesis of no systematic difference between measured distances from \gaia and distances from Type I X-ray bursts. The statistical significances of comparing other methods used in measuring LMXB distances with respect to Gaia distances are reported in \autoref{tab:comparison}. In this table, means and 95\% confidence-intervals of differences are reported, and the p-values are calculated for the null-hypothesis of zero means. According to this table, distances measured using Type I X-ray bursts are the only method which show a systematic difference with new measured distances in this work.

\begin{table}
\caption{
Statistical significance of previous distance measurements of LMXBs comparing to \gaia DR2 distances using Bailer-Jones priors. 
\label{tab:comparison}
} 
\begin{tabular}{lrlll}
\hline
Method & mean &95\%CI  & p-value \\
\hline
\hline
Burst&	$-0.36$&	$[-0.66,-0.06]$& 0.01\\
Photometry&	$0.13$&	$[-0.76,1.02]$& 0.35\\
Radio Parallax&	$-0.19$&	$[-1.22,0.84]$& 0.20\\
Other&	$-0.19$&	$[-0.69,0.31]$& 0.20\\

\hline
\end{tabular}
\end{table}

These results would seem to suggest that Type I X-ray bursts are intrinsically less luminous than predicted by modelling.
This agrees with previous results on systematic biases in distance determination via Type I X-ray bursts. 
\cite{Galloway08a} demonstrated that the choice to assume that the touchdown flux (the flux measured when the expanded photosphere of the neutron star touches down back onto its surface) is either at the Eddington luminosity or sub-Eddington may introduce large systematic uncertainties to distance measurements of X-ray bursting XRBs. 
Studies of bursting sources using the Rossi X-ray Timing Explorer have indicated that a number of these sources are significantly sub-Eddington in their peak fluxes \citep[e.g.,][]{2008ApJS..179..360G}.

As mentioned in \autoref{sec:dist_final_sample}, LMXBs do not follow the same spatial distribution as the stellar distribution assumed by \cite{BailerJones18a}. 
To investigate the effect of the priors on LMXB distances, we also measured the \gaia DR2 distances using the prior developed by \citet{Atri19}, which considers the distribution of LMXBs in the Milky Way based on the work of \citep{Grimm02a}. 
The right panel of \autoref{fig:lmxb_dist} shows the result of our distance comparisons using this prior. 
The most noticeable effect of using the \citet{Atri19} prior is an increase in distances for most of the LMXBs. As a result, distances measured using Type I X-ray bursts are systematically smaller than those measured via \gaia candidate counterparts. 
This is in contradiction with the suggestive results of our analysis using the \citet{BailerJones18a} prior, which indicated that distances based on type I X-ray bursts are overestimated, and thus that bursts only reach 0.5~L$_{\rm Eddington}$. As a model-independent check, we have also done the same analysis without using any prior. In this scenario, the Type I X-ray burst distances are consistent with the null hypothesis of no difference with the \gaia distances, with a $p$-value of $0.19$. This discrepancy highlights the importance of priors when using \gaia\ data for sources with large uncertainties.

Individual objects with particularly large discrepancies between previously-published and \gaia candidate counterpart distances are discussed in Appendix~\ref{sect:apdxB}.

\subsection{Spatial Distribution and Spiral Arms}

To investigate the relationship between XRBs and Galactic structure, we compare the XRB distributions to a model of the spiral arms of the Milky Way. 
\citet{Pettitt2020} show evidence for spiral structure traced by young stars, though it is not clear what the precise physical properties of the spiral structure are.
\citet{Gorski2020} suggest a four arm spiral structure traced by maser-bearing evolved stars.
We use the symmetric arm model of \citet{Vallee08a}. 
This model is analytically defined: the precise shape, symmetries, structure, and extent of the spiral arms of the Galaxy are nontrivial to determine due to our location within the Milky Way.
This symmetric model is fitted to agree with a variety of observations, including dust, H\ion{I} gas, CO gas, and maps of stellar velocities.
This model defines the midpoint of four identical arms phase shifted by 90$^{\circ}$.
We further define the midpoint of interarm regions by shifting the existing arms by 45$^{\circ}$.

 For each XRB, we compute three properties:
 \begin{enumerate}
     \item the two-dimensional distance to the nearest spiral arm for a face-on projection
     \item whether the XRB is leading or trailing its closest spiral arm 
     \item whether the XRB is closer to the midpoint of a spiral arm or the midpoint of an interarm region
 \end{enumerate}

Given that many of the uncertainties for the distances quite large, counts of these quantities depend strongly on the posterior distribution function of the distances.
In order to assess how much these quantities change, we create 10,000 realizations of the distance for each object using the posterior distribution function defined in \cite{BailerJones18a}, and compute the three quantities above for each object in each iteration.

After computing whether each object is closer to an arm or interarm region, whether it is leading or trailing the nearest spiral arm, and the distance to the nearest spiral arm, we calculate the fraction of objects leading/trailing and fraction of objects close to an arm/interarm for each of the 10,000 runs.
Under this construction, since we have effectively partitioned the galaxy into two equally-sized regions (closer to arm/closer to interarm, leading/trailing the nearest spiral arm), we expect the following for the distribution of these fractions:
If the distribution of LMXBs/HMXBs fractions peaks at a value greater than 0.5 for a particular structure (arm/interarm/leading edge/trailing edge), then we interpret that LMXBs/HMXBs as being correlated with that structure.
Conversely, if the distribution peaks at a value less than 0.5, we interpret LMXBs/HMXBs as being anti-correlated with that structure.
 If the distribution peaks at 0.5, we interpret LMXBs/HMXBs as being uncorrelated with that structure. We treat the uncorrelated case as the null hypothesis for LMXBs and HMXBs individually.

\begin{figure*}
\includegraphics[width=0.8\textwidth]{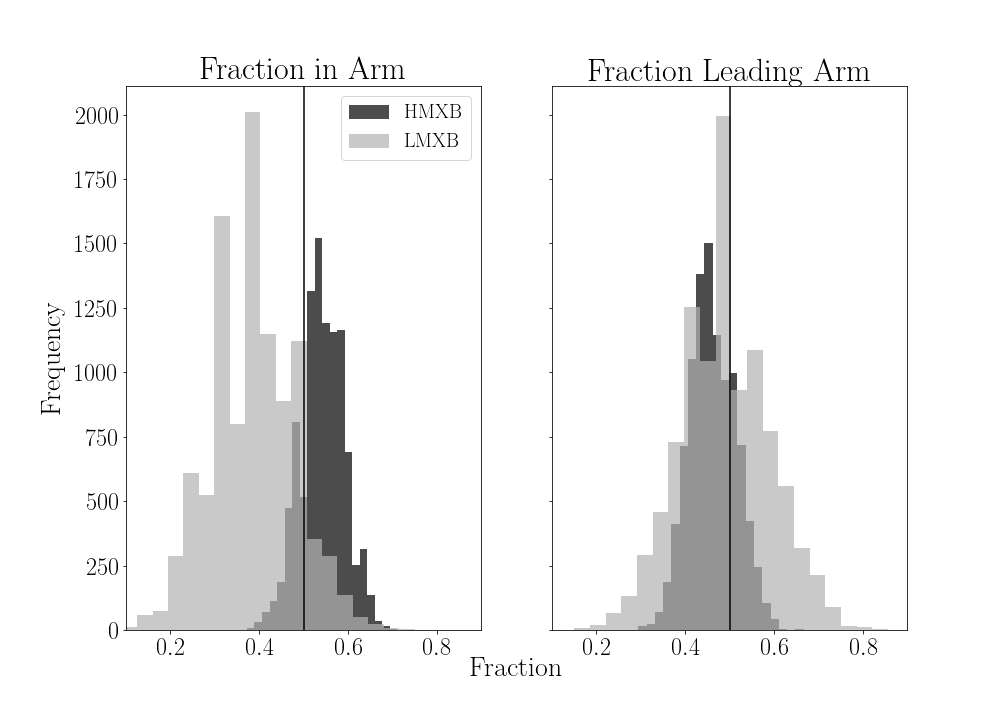}
\caption[Population fraction correlated with spiral arms and leading edges]{Distributions of population fraction correlated with spiral arms and leading edges for 10,000 realizations of the XRBs with \gaia candidate counterparts. The vertical line marks a fraction of 0.5, where the populations would be interpreted as being uncorrelated with the structure.
   }
\label{fig:mc_simulation}
\end{figure*}

In each run, we exclude from the fraction any object that lies at a distance of less than 3.1~kpc from the Galactic centre, classifying them separately as bulge sources.
We choose 3.1~kpc because it is given as the half-length of the bar superimposed on the cartographic plots of \cite{Vallee08a}, and it is noted therein that it becomes difficult to separate the beginnings of the spiral arms from the bar itself at approximately this distance.
In each run, on average two HMXBs and five LMXBs were classified as bulge sources.
The resulting fractions and their uncertainty distributions are plotted in \autoref{fig:mc_simulation}.

\begin{figure}
\includegraphics[width=\columnwidth]{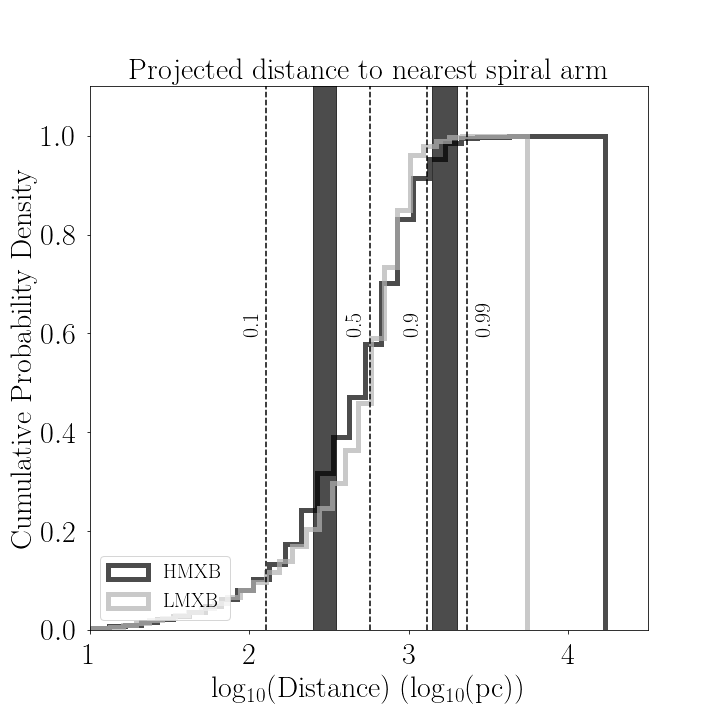}
\caption{
   Cumulative distribution of XRB distances to the nearest spiral arm. We mark the characteristic clustering scales of HMXBs against OB associations and star-forming complexes measured by \citet{Bodaghee12a} and \citet{Coleiro13a} for reference using the vertical grey regions. We also plot the 0.1, 0.5, 0.9, and 0.99 quantiles of the HMXB distribution for comparison.}
\label{fig:dist_to_arms}
\end{figure}

Across the simulation, LMXBs and HMXBs both appear to exhibit a roughly normal distribution in both fractions, though in both the leading/trailing or arm/interarm case, the LMXB distribution possesses a larger spread. 
To compare these measurements to each other and to the null hypothesis (that they are uncorrelated with arms/interarms and leading/trailing spiral arms), we tested these uncertainty distributions for normality. 
Since the interarm/trailing fraction is complementary to the arm/leading fraction, we consider only the arm/leading fractions.
None of the four distributions is considered normal by the D'Agostini $K^{2}$ test or the Anderson-Darling test at $p = 0.05$.
Only the LMXB leading fraction is considered normal by the Jarque-Bera test for $p = 0.05$. 
Since the distributions are not truly normal, we report the fraction measurements and their uncertainty in two ways: first using the standard deviation as the 1$\sigma$ uncertainty, and then reporting the the 95th/5th quantiles as the uncertainty.

Our measurements of the fraction of HMXBs/LMXBs that are correlated with spiral arms/inter-arm regions yields the following results: 
\begin{itemize}
\item Fraction of HMXBs that are closer to a spiral arm: $0.54 \pm 0.05$ (at 1$\sigma$), $0.54^{+0.08}_{-0.08}$ at the 95th and 5th quantiles
\item Fraction of LMXBs that are closer to a spiral arm: $0.39 \pm 0.09$ (at 1$\sigma$), $0.39^{+0.16}_{-0.15}$ at the 95th and 5th quantiles
\item Fraction of HMXBs that are leading the nearest spiral arm: $0.46 \pm 0.05$ (at 1$\sigma$), $0.46^{+0.09}_{-0.09}$ at the 95th and 5th quantiles
\item Fraction of LMXBs that are leading the nearest spiral arm: $0.50 \pm 0.10$ (at 1$\sigma$), $0.49^{+0.16}_{-0.17}$ at the 95th and 5th quantiles
\end{itemize}
We cannot reject the null hypothesis that HMXBs or LMXBs are spatially uncorrelated with spiral arms, at even 1$\sigma$, since the uncertainties overlap with $F_{\rm fraction} = 0.5$.
We cannot reject the null hypothesis for either HMXBs or LMXBs exhibiting no preference leading or trailing their nearest spiral arm.
The LMXB and HMXB fractions also overlap with each other at the $1\sigma$ level.
LMXBs exhibit a mild preference for being found in interarm regions, while HMXBs show only a mild preference for being found in the spiral arms.
LMXBs appear to be uncorrelated with leading or trailing their spiral arm, while at low significance the HMXBs appear to prefer trailing their nearest spiral arm.

In the context of Galactic structure, previous work has shown that HMXBs trace SFR on Galactic scales \citep{Grimm03a}, so it is reasonable to expect they should trace it on resolved scales in some fashion and should exhibit a distinct spatial correlation.
Naively it can be assumed that star formation should happen at the leading edge of a spiral arm where the gas accumulates (see \citealp{Koda12a} for M51 as an illustrative example of star formation and its relation to spiral arm structure).
Taking these assumptions together, HMXBs should be found at the leading edge of spiral arms, and should exhibit a strong preference for spiral arms versus interarm regions.
However, we find only a mild preference for spiral arms: the distribution of fractions for the simulation peaks at 54\% of the HMXBs being closer to an arm than an interarm region, but the wings of the distribution include the uncorrelated and anti-correlated cases.

The lack of strong preference for HMXBs being closely associated with spiral arms could have a number of possible implications:
\begin{itemize}
    \item Star formation does not occur at the leading edge of spiral arms.
    \item The time delay between star formation and HMXB accretion starting manifests itself as a spatial separation between the spiral arm and HMXBs due to the pattern speed of spiral arms.
    \item HMXB natal kicks may be larger than expected.
    \item Our sample is not large enough and does not have sufficiently small distance uncertainties as an ensemble to measure the correlation we expect from first principles.
\end{itemize}
Our HMXB sample comprises only $\sim$50 objects, and the uncertainties are still substantial.
As such, though we can rule out a very strong spatial correlation or anti-correlation between HMXBs and spiral arms (using the Gaia DR2 data specifically), we cannot use our result to distinguish between the scenarios listed above.
Since we are unable to reject the null hypothesis that HMXBs are uncorrelated with spiral arms, our result is consistent with \cite{Bodaghee12a}'s analysis, which found that HMXBs are not spatially correlated with spiral arms.
The scale at which the HMXB/SFR correlation breaks down (if at all) is not well-constrained. 
In nearby galaxies, the X-ray sources are typically studied by considering the integrated properties of the entire population (for example, X-ray luminosity function) and comparing to global parameters of the galaxy. 
Correlating XRBs with galactic structure is challenging since galaxies that are close enough to resolve on the desired scales require many fields in order to encompass the entire galaxy. 
In addition, contamination from X-ray sources in front of or behind the galaxy creates additional difficulties.
\cite{Swartz03a} investigated the relationship between the spiral arms of M81 and its X-ray source population, finding strong correlation between spiral arm position and X-ray source density.
They note that brighter sources tend to be closer to spiral arms, attributable to the brightest and shortest-lived HMXBs being close downstream from their spiral arms.
More recently, \citet{Kuntz16} performed a deep \textit{Chandra} survey of M51.
This study also finds that X-ray sources are concentrated in spiral arms, though the distances to spiral arm midpoints are not presented.
Both studies also found a non-trivial population of supernova remnants contributing to the total X-ray source population.

In contrast to HMXBs, we expect that LMXBs should exhibit no strong preference for spiral arms; they represent (collectively) an older population that is also more strongly perturbed by the strength of its SN kicks \citep{Grimm02a}. 
Since LMXBs can be much older, it is not expected or required that they are still near the spiral arm that formed their progenitor -- there may have been multiple Galactic rotations since the LMXB itself formed.
Additionally, LMXBs' high velocity kicks mean they can be substantially displaced from the star-forming region where they initially formed. 
This process is already required to explain the presence of LMXBs at high Galactic latitudes where they would not be expected to form {\em a priori} due to the low stellar density (see, for example, \citealp{Repetto12a}). 
Consequently, LMXBs as a population should be uncorrelated with spiral arms since their distribution would be unperturbed by either the presence or absence of spiral arms. 
This makes our result, which shows LMXBs anti-correlated with spiral arms (though at low significance), difficult to explain.

We also computed the distribution of distances to the nearest spiral arm across all the simulations, shown in \autoref{fig:dist_to_arms}, in order to compare with previous works that measured the distances to OB associations and star-forming complexes for HMXBs \citep{Bodaghee12a,Coleiro13a}.
In these works, clustering distances between HMXBs and SFCs/OB associations were inferred from the critical points of the cumulative distribution of the distances to the nearest SFC/OB association.
As discussed in Section \ref{xraybinaries}, distances to OB associations and SF complexes are distinct from distances to the spiral arms themselves, and as such we might not expect HMXBs to have the same clustering distance to the spiral arm.
The distribution of HMXB distances to the nearest spiral arm that we measure does not show a strong preference for the clustering sizes measured for OB associations or SF complexes in previous works, though we note that the \cite{Vallee08a} model does not fit spiral arms to either of these structures.
The 0.1, 0.5, 0.9, and 0.99 quantiles of the HMXB distribution to be at 127, 570, 1296, and 2340~pc, respectively.
For the LMXB distribution, the 0.1, 0.5, 0.9, and 0.99 quantiles of the LMXB distribution are at 130, 610, 1090, and 1780~pc, respectively. 

Given the substantial width of these distributions, it is difficult to determine a characteristic separation from the spiral arms.
The interarm separation of a few kpc as set by the symmetric arms model means that, by construction, it is difficult to have an XRB more than a few kpc away from a spiral arm in face-on projection.
Further, we have chosen to model the galaxy using a symmetric model fitted to observables in the Milky Way, which is a simple albeit potentially unrealistic choice.
The primary advantage of this model is that it permits us to easily define inter-arm and arm regions for analysis of the locations of XRBs.
In reality, the number of arms and the symmetry (e.g., are the four arms symmetric with each other or are there major/minor axes?) of these arms in the Milky Way is difficult to characterize (see \citealt{Vallee17a} and references therein), and discussion exists about which tracers to use and how far to project the model based on nearby observables.
Future attempts to characterize the relationship between the Galaxy's spiral arms and its XRB population would be improved by the use of a model that relaxes the symmetry constraint.

\section{Conclusions}

\begin{itemize}
    \item We have assembled the largest sample of Galactic X-ray binaries whose distances have all been measured using the same method, and hence have the same systematics and uniform presumed biases.
    \item Comparing XRB distances measured by \gaia (using the Bailer-Jones prior) to previous methods shows that measuring distances using Type I X-ray bursts appears to systematically overestimate distance.
    This suggests that assumptions about X-ray bursts, namely that bursting neutron stars consistently reach the Eddington luminosity, may need to be modified to use X-ray bursts as a distance estimator. This effect is prior-dependent, as choosing a different prior, such as the one in \cite{Atri19}, can cause burst distances to be systematically lower than those from \gaia DR2.
    \item We have compared the positions of XRBs to the locations of the midpoints of spiral arms in the Milky Way. Galactic HMXBs in our sample show only a modest preference for being spatially co-located with spiral arms versus interarm regions, and show only a modest preference for being on the leading edge of spiral arms. This suggests that the delay time between star formation and HMXB formation/accretion beginning manifests itself observationally as a spatial separation between HMXBs and spiral arms due to the pattern speed of spiral arm rotation. Other possible explanations for this effect are scattering due to natal HMXB kicks or the possibility of star formation occurring closer to the midpoint of the arm than the leading edge.
    \item We further find that HMXB distances to the nearest spiral arm do not show a strong preference for the clustering sizes previously observed for OB associations or SF complexes.
    \item We find that LMXBs are very weakly anti-correlated with spiral arms. This disagrees with the expectation that LMXBs should be uncorrelated with spiral arms, though we note that the significance of this result is low.
    
\end{itemize}

A main source of uncertainty in our analysis is the low number of XRBs with \gaia counterparts.
Further releases of \gaia will hopefully yield additional \gaia candidate counterparts for Galactic XRBs, particularly for the intrinsically optically fainter LMXBs.
For objects with identified \gaia candidate counterparts, smaller distance uncertainties are expected from the improved baseline in DR3 and subsequent releases.
The  small sample size from the Liu catalogues is another limitation of our analysis.
The \textit{Chandra} Source Catalog \citep{evans2010} provides an excellent foundation for studying the Galactic X-ray sky in the \textit{Chandra} era, but at present it has not been data-mined to make a Milky Way-specific catalogue as a potential successor to the Liu catalogues. 
Our knowledge of the Galactic XRB source population can be improved through future all-sky surveys, such as with the newly-launched eROSITA mission \citep{Merloni12a}.
This mission, designed as a successor to the ROSAT mission, will survey the sky at approximately 20 times the sensitivity of ROSAT in soft X-rays (0.5--2.0~keV), while providing the first imaging survey of the sky in hard X-rays (2--10~keV).
The on-axis angular resolution of this telescope is expected to be comparable to that of \textit{XMM-Newton}. 
An improved all-sky survey will allow us to find \gaia counterparts to an X-ray catalogue that is more up-to-date and is has more uniform systematics, enhancing our understanding of how XRB positions correlate with Galactic structure. 

\section*{Acknowledgements}

We thank the referee, P. Gandhi, for constructive and helpful reports.
This work has made use of data from the European Space Agency (ESA) mission
{\it Gaia} (\url{https://www.cosmos.esa.int/gaia}), processed by the {\it Gaia}
Data Processing and Analysis Consortium (DPAC,
\url{https://www.cosmos.esa.int/web/gaia/dpac/consortium}). Funding for the DPAC 
has been provided by national institutions, in particular the institutions
participating in the {\it Gaia} Multilateral Agreement.
R. M. A. acknowledges support from an NSERC CGS-D scholarship. 
P. B. acknowledges support from an NSERC Discovery Grant and the hospitality of the Rotman Institute for Philosophy.
We thank E. Cackett, S. Gallagher, and T.A.A. Sigut for helpful discussions.

\section*{Data availability}

The data underlying this article are available in the article and in its online supplementary material.

\bibliographystyle{mnras}
\bibliography{xrb_all} 

\begin{thebibliography}{}
\makeatletter
\relax
\def\mn@urlcharsother{\let\do\@makeother \do\$\do\&\do\#\do\^\do\_\do\%\do\~}
\def\mn@doi{\begingroup\mn@urlcharsother \@ifnextchar [ {\mn@doi@}
  {\mn@doi@[]}}
\def\mn@doi@[#1]#2{\def\@tempa{#1}\ifx\@tempa\@empty \href
  {http://dx.doi.org/#2} {doi:#2}\else \href {http://dx.doi.org/#2} {#1}\fi
  \endgroup}
\def\mn@eprint#1#2{\mn@eprint@#1:#2::\@nil}
\def\mn@eprint@arXiv#1{\href {http://arxiv.org/abs/#1} {{\tt arXiv:#1}}}
\def\mn@eprint@dblp#1{\href {http://dblp.uni-trier.de/rec/bibtex/#1.xml}
  {dblp:#1}}
\def\mn@eprint@#1:#2:#3:#4\@nil{\def\@tempa {#1}\def\@tempb {#2}\def\@tempc
  {#3}\ifx \@tempc \@empty \let \@tempc \@tempb \let \@tempb \@tempa \fi \ifx
  \@tempb \@empty \def\@tempb {arXiv}\fi \@ifundefined
  {mn@eprint@\@tempb}{\@tempb:\@tempc}{\expandafter \expandafter \csname
  mn@eprint@\@tempb\endcsname \expandafter{\@tempc}}}

\bibitem[\protect\citeauthoryear{{Arenou} et~al.}{{Arenou}
  et~al.}{2018}]{arenou2018}
{Arenou} F.,  et~al., 2018, \mn@doi [\aap] {10.1051/0004-6361/201833234}, \href
  {https://ui.adsabs.harvard.edu/abs/2018A&A...616A..17A} {616, A17}

\bibitem[\protect\citeauthoryear{{Atri} et~al.,}{{Atri} et~al.}{2019}]{Atri19}
{Atri} P.,  et~al., 2019, \mn@doi [\mnras] {10.1093/mnras/stz2335}, \href
  {https://ui.adsabs.harvard.edu/abs/2019MNRAS.489.3116A} {489, 3116}

\bibitem[\protect\citeauthoryear{{Augusteijn}, {van der Hooft}, {de Jong}, {van
  Kerkwijk}  \& {van Paradijs}}{{Augusteijn} et~al.}{1998}]{1998AA...332..561A}
{Augusteijn} T.,  {van der Hooft} F.,  {de Jong} J.~A.,  {van Kerkwijk} M.~H.,
   {van Paradijs} J.,  1998, \aap, \href
  {https://ui.adsabs.harvard.edu/#abs/1998A&A...332..561A} {332, 561}

\bibitem[\protect\citeauthoryear{{Bahramian}, {Heinke}, {Degenaar}, {Chomiuk},
  {Wijnands}, {Strader}, {Ho}  \& {Pooley}}{{Bahramian}
  et~al.}{2015}]{Bahramian2015}
{Bahramian} A.,  {Heinke} C.~O.,  {Degenaar} N.,  {Chomiuk} L.,  {Wijnands} R.,
   {Strader} J.,  {Ho} W. C.~G.,   {Pooley} D.,  2015, \mn@doi [\mnras]
  {10.1093/mnras/stv1585}, \href
  {https://ui.adsabs.harvard.edu/abs/2015MNRAS.452.3475B} {452, 3475}

\bibitem[\protect\citeauthoryear{{Bailer-Jones}, {Rybizki}, {Fouesneau},
  {Mantelet}  \& {Andrae}}{{Bailer-Jones} et~al.}{2018}]{BailerJones18a}
{Bailer-Jones} C.~A.~L.,  {Rybizki} J.,  {Fouesneau} M.,  {Mantelet} G.,
  {Andrae} R.,  2018, \mn@doi [\aj] {10.3847/1538-3881/aacb21}, 156, 58

\bibitem[\protect\citeauthoryear{{Bamba}, {Yokogawa}, {Ueno}, {Koyama}  \&
  {Yamauchi}}{{Bamba} et~al.}{2001}]{2001PASJ...53.1179B}
{Bamba} A.,  {Yokogawa} J.,  {Ueno} M.,  {Koyama} K.,   {Yamauchi} S.,  2001,
  \mn@doi [\pasj] {10.1093/pasj/53.6.1179}, \href
  {https://ui.adsabs.harvard.edu/#abs/2001PASJ...53.1179B} {53, 1179}

\bibitem[\protect\citeauthoryear{{Basinska}, {Lewin}, {Sztajno}, {Cominsky}  \&
  {Marshall}}{{Basinska} et~al.}{1984}]{Basinska84a}
{Basinska} E.~M.,  {Lewin} W.~H.~G.,  {Sztajno} M.,  {Cominsky} L.~R.,
  {Marshall} F.~J.,  1984, \mn@doi [\apj] {10.1086/162103}, \href
  {http://adsabs.harvard.edu/abs/1984ApJ...281..337B} {281, 337}

\bibitem[\protect\citeauthoryear{{Bassa}, {Jonker}, {in't Zand}  \&
  {Verbunt}}{{Bassa} et~al.}{2006}]{2006AA...446L..17B}
{Bassa} C.~G.,  {Jonker} P.~G.,  {in't Zand} J.~J.~M.,   {Verbunt} F.,  2006,
  \mn@doi [\aap] {10.1051/0004-6361:200500229}, \href
  {https://ui.adsabs.harvard.edu/#abs/2006A&A...446L..17B} {446, L17}

\bibitem[\protect\citeauthoryear{{Baumgartner}, {Tueller}, {Markwardt},
  {Skinner}, {Barthelmy}, {Mushotzky}, {Evans}  \& {Gehrels}}{{Baumgartner}
  et~al.}{2013}]{Baumgartner2013}
{Baumgartner} W.~H.,  {Tueller} J.,  {Markwardt} C.~B.,  {Skinner} G.~K.,
  {Barthelmy} S.,  {Mushotzky} R.~F.,  {Evans} P.~A.,   {Gehrels} N.,  2013,
  \mn@doi [\apjs] {10.1088/0067-0049/207/2/19}, 207, 19

\bibitem[\protect\citeauthoryear{{Belokurov} et~al.,}{{Belokurov}
  et~al.}{2020}]{Belokurov2020}
{Belokurov} V.,  et~al., 2020, \mn@doi [\mnras] {10.1093/mnras/staa1522}, \href
  {https://ui.adsabs.harvard.edu/abs/2020MNRAS.496.1922B} {496, 1922}

\bibitem[\protect\citeauthoryear{{Berghoefer}, {Schmitt}  \&
  {Cassinelli}}{{Berghoefer} et~al.}{1996}]{Berghoefer96a}
{Berghoefer} T.~W.,  {Schmitt} J.~H.~M.~M.,   {Cassinelli} J.~P.,  1996,
  Astronomy and Astrophysics Supplement Series, \href
  {https://ui.adsabs.harvard.edu/#abs/1996A&AS..118..481B} {118, 481}

\bibitem[\protect\citeauthoryear{{Bhattacharyya}}{{Bhattacharyya}}{2010}]{Bhattacharyya10a}
{Bhattacharyya} S.,  2010, \mn@doi [Advances in Space Research]
  {10.1016/j.asr.2010.01.010}, \href
  {http://adsabs.harvard.edu/abs/2010AdSpR..45..949B} {45, 949}

\bibitem[\protect\citeauthoryear{{Blay}, {Negueruela}, {Reig}, {Coe}, {Corbet},
  {Fabregat}  \& {Tarasov}}{{Blay} et~al.}{2006}]{2006AA...446.1095B}
{Blay} P.,  {Negueruela} I.,  {Reig} P.,  {Coe} M.~J.,  {Corbet} R.~H.~D.,
  {Fabregat} J.,   {Tarasov} A.~E.,  2006, \mn@doi [\aap]
  {10.1051/0004-6361:20053951}, \href
  {https://ui.adsabs.harvard.edu/#abs/2006A&A...446.1095B} {446, 1095}

\bibitem[\protect\citeauthoryear{{Bodaghee}, {Tomsick}, {Rodriguez}  \&
  {James}}{{Bodaghee} et~al.}{2012}]{Bodaghee12a}
{Bodaghee} A.,  {Tomsick} J.~A.,  {Rodriguez} J.,   {James} J.~B.,  2012,
  \mn@doi [\apj] {10.1088/0004-637X/744/2/108}, \href
  {http://adsabs.harvard.edu/abs/2012ApJ...744..108B} {744, 108}

\bibitem[\protect\citeauthoryear{{Boroson}, {Kim}  \& {Fabbiano}}{{Boroson}
  et~al.}{2011}]{Boroson11a}
{Boroson} B.,  {Kim} D.-W.,   {Fabbiano} G.,  2011, \mn@doi [\apj]
  {10.1088/0004-637X/729/1/12}, \href
  {http://adsabs.harvard.edu/abs/2011ApJ...729...12B} {729, 12}

\bibitem[\protect\citeauthoryear{{Boubert}, {Guillochon}, {Hawkins},
  {Ginsburg}, {Evans}  \& {Strader}}{{Boubert} et~al.}{2018}]{Boubert18a}
{Boubert} D.,  {Guillochon} J.,  {Hawkins} K.,  {Ginsburg} I.,  {Evans} N.~W.,
   {Strader} J.,  2018, \mn@doi [\mnras] {10.1093/mnras/sty1601}, \href
  {http://adsabs.harvard.edu/abs/2018MNRAS.479.2789B} {479, 2789}

\bibitem[\protect\citeauthoryear{{Bozzo} et~al.,}{{Bozzo}
  et~al.}{2018}]{Bozzo18a}
{Bozzo} E.,  et~al., 2018, \mn@doi [\aap] {10.1051/0004-6361/201832588}, \href
  {http://adsabs.harvard.edu/abs/2018A\%26A...613A..22B} {613, A22}

\bibitem[\protect\citeauthoryear{{Bradshaw}, {Fomalont}  \&
  {Geldzahler}}{{Bradshaw} et~al.}{1999}]{1999ApJ...512L.121B}
{Bradshaw} C.~F.,  {Fomalont} E.~B.,   {Geldzahler} B.~J.,  1999, \mn@doi
  [\apj] {10.1086/311889}, \href
  {https://ui.adsabs.harvard.edu/#abs/1999ApJ...512L.121B} {512, L121}

\bibitem[\protect\citeauthoryear{{Brandt}, {Castro-Tirado}, {Lund}, {Dremin},
  {Lapshov}  \& {Sunyaev}}{{Brandt} et~al.}{1992}]{1992AA...262L..15B}
{Brandt} S.,  {Castro-Tirado} A.~J.,  {Lund} N.,  {Dremin} V.,  {Lapshov} I.,
  {Sunyaev} R.,  1992, \aap, \href
  {https://ui.adsabs.harvard.edu/#abs/1992A&A...262L..15B} {262, L15}

\bibitem[\protect\citeauthoryear{{Brown}, {Blaauw}, {Hoogerwerf}, {de Bruijne}
  \& {de Zeeuw}}{{Brown} et~al.}{1999}]{Brown99a}
{Brown} A.~G.~A.,  {Blaauw} A.,  {Hoogerwerf} R.,  {de Bruijne} J.~H.~J.,   {de
  Zeeuw} P.~T.,  1999, in {Lada} C.~J.,  {Kylafis} N.~D.,  eds,  NATO Advanced
  Science Institutes (ASI) Series C Vol. 540, NATO Advanced Science Institutes
  (ASI) Series C. p.~411 (\mn@eprint {} {astro-ph/9902234})

\bibitem[\protect\citeauthoryear{{Butters}, {Norton}, {Mukai}  \&
  {Tomsick}}{{Butters} et~al.}{2011}]{Butters2011}
{Butters} O.~W.,  {Norton} A.~J.,  {Mukai} K.,   {Tomsick} J.~A.,  2011,
  \mn@doi [\aap] {10.1051/0004-6361/201015848}, 526, A77

\bibitem[\protect\citeauthoryear{{Cadolle Bel} et~al.,}{{Cadolle Bel}
  et~al.}{2007}]{2007ApJ...659..549C}
{Cadolle Bel} M.,  et~al., 2007, \mn@doi [\apj] {10.1086/512004}, \href
  {https://ui.adsabs.harvard.edu/#abs/2007ApJ...659..549C} {659, 549}

\bibitem[\protect\citeauthoryear{{Cantrell} et~al.,}{{Cantrell}
  et~al.}{2010}]{2010ApJ...710.1127C}
{Cantrell} A.~G.,  et~al., 2010, \mn@doi [\apj] {10.1088/0004-637X/710/2/1127},
  \href {https://ui.adsabs.harvard.edu/abs/2010ApJ...710.1127C} {710, 1127}

\bibitem[\protect\citeauthoryear{{Casares}, {Rib{\'o}}, {Ribas}, {Paredes},
  {Mart{\'\i}}  \& {Herrero}}{{Casares} et~al.}{2005}]{2005MNRAS.364..899C}
{Casares} J.,  {Rib{\'o}} M.,  {Ribas} I.,  {Paredes} J.~M.,  {Mart{\'\i}} J.,
   {Herrero} A.,  2005, \mn@doi [\mnras] {10.1111/j.1365-2966.2005.09617.x},
  \href {https://ui.adsabs.harvard.edu/#abs/2005MNRAS.364..899C} {364, 899}

\bibitem[\protect\citeauthoryear{{Casares}, {Jonker}  \& {Israelian}}{{Casares}
  et~al.}{2017}]{Casares17a}
{Casares} J.,  {Jonker} P.~G.,   {Israelian} G.,  2017, in {Alsabti} A.~W.,
  {Murdin} P.,  eds, , Handbook of Supernovae.
Springer International Publishing AG, p.~1499,
  \mn@doi{10.1007/978-3-319-21846-5\_111}

\bibitem[\protect\citeauthoryear{{Chakrabarty} \& {Roche}}{{Chakrabarty} \&
  {Roche}}{1997}]{1997ApJ...489..254C}
{Chakrabarty} D.,  {Roche} P.,  1997, \mn@doi [\apj] {10.1086/304779}, \href
  {https://ui.adsabs.harvard.edu/#abs/1997ApJ...489..254C} {489, 254}

\bibitem[\protect\citeauthoryear{{Chan} \& {Bovy}}{{Chan} \&
  {Bovy}}{2020}]{chan2020}
{Chan} V.~C.,  {Bovy} J.,  2020, \mn@doi [\mnras] {10.1093/mnras/staa571},
  \href {https://ui.adsabs.harvard.edu/abs/2020MNRAS.493.4367C} {493, 4367}

\bibitem[\protect\citeauthoryear{{Chernyakova}, {Lutovinov}, {Rodr{\'\i}guez}
  \& {Revnivtsev}}{{Chernyakova} et~al.}{2005}]{2005MNRAS.364..455C}
{Chernyakova} M.,  {Lutovinov} A.,  {Rodr{\'\i}guez} J.,   {Revnivtsev} M.,
  2005, \mn@doi [\mnras] {10.1111/j.1365-2966.2005.09548.x}, \href
  {https://ui.adsabs.harvard.edu/#abs/2005MNRAS.364..455C} {364, 455}

\bibitem[\protect\citeauthoryear{{Chevalier} \& {Ilovaisky}}{{Chevalier} \&
  {Ilovaisky}}{1998}]{1998AA...330..201C}
{Chevalier} C.,  {Ilovaisky} S.~A.,  1998, \aap, \href
  {https://ui.adsabs.harvard.edu/#abs/1998A&A...330..201C} {330, 201}

\bibitem[\protect\citeauthoryear{{Clark}}{{Clark}}{1975}]{Clarke75a}
{Clark} G.~W.,  1975, \mn@doi [\apjl] {10.1086/181869}, \href
  {http://adsabs.harvard.edu/abs/1975ApJ...199L.143C} {199, L143}

\bibitem[\protect\citeauthoryear{{Clark}}{{Clark}}{2004}]{2004ApJ...610..956C}
{Clark} G.~W.,  2004, \mn@doi [\apj] {10.1086/421764}, \href
  {https://ui.adsabs.harvard.edu/#abs/2004ApJ...610..956C} {610, 956}

\bibitem[\protect\citeauthoryear{{Coe}, {Everall}, {Norton}, {Roche}, {Unger},
  {Fabregat}, {Reglero}  \& {Grunsfeld}}{{Coe}
  et~al.}{1993}]{1993MNRAS.261..599C}
{Coe} M.~J.,  {Everall} C.,  {Norton} A.~J.,  {Roche} P.,  {Unger} S.~J.,
  {Fabregat} J.,  {Reglero} V.,   {Grunsfeld} J.~M.,  1993, \mn@doi [\mnras]
  {10.1093/mnras/261.3.599}, \href
  {https://ui.adsabs.harvard.edu/#abs/1993MNRAS.261..599C} {261, 599}

\bibitem[\protect\citeauthoryear{{Coe} et~al.,}{{Coe}
  et~al.}{1994}]{1994MNRAS.270L..57C}
{Coe} M.~J.,  et~al., 1994, \mn@doi [\mnras] {10.1093/mnras/270.1.L57}, \href
  {https://ui.adsabs.harvard.edu/#abs/1994MNRAS.270L..57C} {270, L57}

\bibitem[\protect\citeauthoryear{{Coleiro} \& {Chaty}}{{Coleiro} \&
  {Chaty}}{2013}]{Coleiro13a}
{Coleiro} A.,  {Chaty} S.,  2013, \mn@doi [\apj] {10.1088/0004-637X/764/2/185},
  \href {http://adsabs.harvard.edu/abs/2013ApJ...764..185C} {764, 185}

\bibitem[\protect\citeauthoryear{{Coley}, {Corbet}  \& {Krimm}}{{Coley}
  et~al.}{2015}]{2015ApJ...808..140C}
{Coley} J.~B.,  {Corbet} R. H.~D.,   {Krimm} H.~A.,  2015, \mn@doi [\apj]
  {10.1088/0004-637X/808/2/140}, \href
  {https://ui.adsabs.harvard.edu/abs/2015ApJ...808..140C} {808, 140}

\bibitem[\protect\citeauthoryear{{Corbet} \& {Mason}}{{Corbet} \&
  {Mason}}{1984}]{1984AA...131..385C}
{Corbet} R.~H.~D.,  {Mason} K.~O.,  1984, \aap, \href
  {https://ui.adsabs.harvard.edu/#abs/1984A&A...131..385C} {131, 385}

\bibitem[\protect\citeauthoryear{{Corral-Santana}, {Casares},
  {Mu{\~n}oz-Darias}, {Bauer}, {Mart{\'\i}nez-Pais}  \&
  {Russell}}{{Corral-Santana} et~al.}{2016}]{blackcat2016}
{Corral-Santana} J.~M.,  {Casares} J.,  {Mu{\~n}oz-Darias} T.,  {Bauer} F.~E.,
  {Mart{\'\i}nez-Pais} I.~G.,   {Russell} D.~M.,  2016, \mn@doi [\aap]
  {10.1051/0004-6361/201527130}, 587, A61

\bibitem[\protect\citeauthoryear{{Cowley} \& {Schmidtke}}{{Cowley} \&
  {Schmidtke}}{1990}]{1990AJ.....99..678C}
{Cowley} A.~P.,  {Schmidtke} P.~C.,  1990, \mn@doi [\aj] {10.1086/115363},
  \href {https://ui.adsabs.harvard.edu/#abs/1990AJ.....99..678C} {99, 678}

\bibitem[\protect\citeauthoryear{{Dabringhausen}, {Kroupa}, {Pflamm-Altenburg}
  \& {Mieske}}{{Dabringhausen} et~al.}{2012}]{Dabringhausen12a}
{Dabringhausen} J.,  {Kroupa} P.,  {Pflamm-Altenburg} J.,   {Mieske} S.,  2012,
  \mn@doi [\apj] {10.1088/0004-637X/747/1/72}, \href
  {http://adsabs.harvard.edu/abs/2012ApJ...747...72D} {747, 72}

\bibitem[\protect\citeauthoryear{{Degenaar}, {Wijnands}, {Cackett}, {Homan},
  {in't Zand}, {Kuulkers}, {Maccarone}  \& {van der Klis}}{{Degenaar}
  et~al.}{2012}]{Degenaar12a}
{Degenaar} N.,  {Wijnands} R.,  {Cackett} E.~M.,  {Homan} J.,  {in't Zand}
  J.~J.~M.,  {Kuulkers} E.,  {Maccarone} T.~J.,   {van der Klis} M.,  2012,
  \mn@doi [\aap] {10.1051/0004-6361/201219470}, \href
  {http://adsabs.harvard.edu/abs/2012A\%26A...545A..49D} {545, A49}

\bibitem[\protect\citeauthoryear{{Dhawan}, {Mirabel}, {Rib{\'o}}  \&
  {Rodrigues}}{{Dhawan} et~al.}{2007}]{Dhawan07a}
{Dhawan} V.,  {Mirabel} I.~F.,  {Rib{\'o}} M.,   {Rodrigues} I.,  2007, \mn@doi
  [\apj] {10.1086/520111}, \href
  {http://adsabs.harvard.edu/abs/2007ApJ...668..430D} {668, 430}

\bibitem[\protect\citeauthoryear{{Erkal}, {Boubert}, {Gualandris}, {Evans}  \&
  {Antonini}}{{Erkal} et~al.}{2019}]{Erkal19}
{Erkal} D.,  {Boubert} D.,  {Gualandris} A.,  {Evans} N.~W.,   {Antonini} F.,
  2019, \mn@doi [\mnras] {10.1093/mnras/sty2674}, 483, 2007

\bibitem[\protect\citeauthoryear{{Evans} et~al.}{{Evans}
  et~al.}{2010}]{evans2010}
{Evans} I.~N.,  et~al., 2010, \mn@doi [\apjs] {10.1088/0067-0049/189/1/37},
  189, 37

\bibitem[\protect\citeauthoryear{{Filliatre} \& {Chaty}}{{Filliatre} \&
  {Chaty}}{2004}]{2004ApJ...616..469F}
{Filliatre} P.,  {Chaty} S.,  2004, \mn@doi [\apj] {10.1086/424869}, \href
  {https://ui.adsabs.harvard.edu/#abs/2004ApJ...616..469F} {616, 469}

\bibitem[\protect\citeauthoryear{{Foellmi}, {Depagne}, {Dall}  \&
  {Mirabel}}{{Foellmi} et~al.}{2006}]{2006A&A...457..249F}
{Foellmi} C.,  {Depagne} E.,  {Dall} T.~H.,   {Mirabel} I.~F.,  2006, \mn@doi
  [\aap] {10.1051/0004-6361:20054686}, \href
  {https://ui.adsabs.harvard.edu/#abs/2006A&A...457..249F} {457, 249}

\bibitem[\protect\citeauthoryear{{Foight}, {G{\"u}ver}, {{\"O}zel}  \&
  {Slane}}{{Foight} et~al.}{2016}]{Foight2016}
{Foight} D.~R.,  {G{\"u}ver} T.,  {{\"O}zel} F.,   {Slane} P.~O.,  2016,
  \mn@doi [\apj] {10.3847/0004-637X/826/1/66}, \href
  {https://ui.adsabs.harvard.edu/abs/2016ApJ...826...66F} {826, 66}

\bibitem[\protect\citeauthoryear{{Fortin}, {Chaty}  \& {Sander}}{{Fortin}
  et~al.}{2020}]{2020ApJ...894...86F}
{Fortin} F.,  {Chaty} S.,   {Sander} A.,  2020, \mn@doi [\apj]
  {10.3847/1538-4357/ab881c}, \href
  {https://ui.adsabs.harvard.edu/abs/2020ApJ...894...86F} {894, 86}

\bibitem[\protect\citeauthoryear{{Gaia Collaboration} et~al.,}{{Gaia
  Collaboration} et~al.}{2016a}]{gaia2016a}
{Gaia Collaboration} et~al., 2016a, \mn@doi [\aap]
  {10.1051/0004-6361/201629272}, \href
  {http://adsabs.harvard.edu/abs/2016A\%26A...595A...1G} {595, A1}

\bibitem[\protect\citeauthoryear{{Gaia Collaboration} et~al.,}{{Gaia
  Collaboration} et~al.}{2016b}]{gaiaDR1}
{Gaia Collaboration} et~al., 2016b, \mn@doi [\aap]
  {10.1051/0004-6361/201629512}, \href
  {http://adsabs.harvard.edu/abs/2016A\%26A...595A...2G} {595, A2}

\bibitem[\protect\citeauthoryear{{Gaia Collaboration} et~al.,}{{Gaia
  Collaboration} et~al.}{2018}]{gaiaDR2}
{Gaia Collaboration} et~al., 2018, \mn@doi [\aap]
  {10.1051/0004-6361/201833051}, \href
  {http://adsabs.harvard.edu/abs/2018A\%26A...616A...1G} {616, A1}

\bibitem[\protect\citeauthoryear{{Gaia Collaboration}, Brown, Vallenari,
  Prusti, de Bruijne, Babusiaux  \& Biermann}{{Gaia Collaboration}
  et~al.}{2020}]{gaiaEDR3}
{Gaia Collaboration} Brown A. G.~A.,  Vallenari A.,  Prusti T.,  de Bruijne J.
  H.~J.,  Babusiaux C.,   Biermann M.,  2020, \aap, in press

\bibitem[\protect\citeauthoryear{{Galloway}, {Psaltis}, {Chakrabarty}  \&
  {Muno}}{{Galloway} et~al.}{2003}]{Galloway03a}
{Galloway} D.~K.,  {Psaltis} D.,  {Chakrabarty} D.,   {Muno} M.~P.,  2003,
  \mn@doi [\apj] {10.1086/375049}, \href
  {http://adsabs.harvard.edu/abs/2003ApJ...590..999G} {590, 999}

\bibitem[\protect\citeauthoryear{{Galloway}, {Muno}, {Hartman}, {Psaltis}  \&
  {Chakrabarty}}{{Galloway} et~al.}{2008a}]{2008ApJS..179..360G}
{Galloway} D.~K.,  {Muno} M.~P.,  {Hartman} J.~M.,  {Psaltis} D.,
  {Chakrabarty} D.,  2008a, \mn@doi [\apjs] {10.1086/592044}, \href
  {https://ui.adsabs.harvard.edu/#abs/2008ApJS..179..360G} {179, 360}

\bibitem[\protect\citeauthoryear{{Galloway}, {{\"O}zel}  \&
  {Psaltis}}{{Galloway} et~al.}{2008b}]{Galloway08a}
{Galloway} D.~K.,  {{\"O}zel} F.,   {Psaltis} D.,  2008b, \mn@doi [\mnras]
  {10.1111/j.1365-2966.2008.13219.x}, \href
  {http://adsabs.harvard.edu/abs/2008MNRAS.387..268G} {387, 268}

\bibitem[\protect\citeauthoryear{{Gandhi}, {Rao}, {Johnson}, {Paice}  \&
  {Maccarone}}{{Gandhi} et~al.}{2019}]{Gandhi19}
{Gandhi} P.,  {Rao} A.,  {Johnson} M. A.~C.,  {Paice} J.~A.,   {Maccarone}
  T.~J.,  2019, \mn@doi [\mnras] {10.1093/mnras/stz438}, 485, 2642

\bibitem[\protect\citeauthoryear{{Gelino}}{{Gelino}}{2001}]{2001PhDT..........G}
{Gelino} D.~M.,  2001, PhD thesis, Center for Astrophysics and Space Sciences,
  University of California, San Diego

\bibitem[\protect\citeauthoryear{{Gelino} \& {Harrison}}{{Gelino} \&
  {Harrison}}{2003}]{2003ApJ...599.1254G}
{Gelino} D.~M.,  {Harrison} T.~E.,  2003, \mn@doi [\apj] {10.1086/379311},
  \href {https://ui.adsabs.harvard.edu/abs/2003ApJ...599.1254G} {599, 1254}

\bibitem[\protect\citeauthoryear{{Gilfanov}}{{Gilfanov}}{2004}]{Gilfanov04a}
{Gilfanov} M.,  2004, \mn@doi [\mnras] {10.1111/j.1365-2966.2004.07473.x},
  \href {http://adsabs.harvard.edu/abs/2004MNRAS.349..146G} {349, 146}

\bibitem[\protect\citeauthoryear{{Gonz{\'a}lez Hern{\'a}ndez}, {Rebolo},
  {Pe{\~n}arrubia}, {Casares}  \& {Israelian}}{{Gonz{\'a}lez Hern{\'a}ndez}
  et~al.}{2005}]{GonzalezHernandez05a}
{Gonz{\'a}lez Hern{\'a}ndez} J.~I.,  {Rebolo} R.,  {Pe{\~n}arrubia} J.,
  {Casares} J.,   {Israelian} G.,  2005, \mn@doi [\aap]
  {10.1051/0004-6361:20042453}, \href
  {http://adsabs.harvard.edu/abs/2005A\%26A...435.1185G} {435, 1185}

\bibitem[\protect\citeauthoryear{{Gorski} \& {Barmby}}{{Gorski} \&
  {Barmby}}{2020}]{Gorski2020}
{Gorski} M.~D.,  {Barmby} P.,  2020, \mn@doi [\mnras] {10.1093/mnras/staa1187},
  \href {https://ui.adsabs.harvard.edu/abs/2020MNRAS.495..726G} {495, 726}

\bibitem[\protect\citeauthoryear{{Green}, {Schlafly}, {Zucker}, {Speagle}  \&
  {Finkbeiner}}{{Green} et~al.}{2019}]{2019ApJ...887...93G}
{Green} G.~M.,  {Schlafly} E.,  {Zucker} C.,  {Speagle} J.~S.,   {Finkbeiner}
  D.,  2019, \mn@doi [\apj] {10.3847/1538-4357/ab5362}, \href
  {https://ui.adsabs.harvard.edu/abs/2019ApJ...887...93G} {887, 93}

\bibitem[\protect\citeauthoryear{{Grillo}, {Sciortino}, {Micela}, {Vaiana}  \&
  {Harnden}}{{Grillo} et~al.}{1992}]{1992ApJS...81..795G}
{Grillo} F.,  {Sciortino} S.,  {Micela} G.,  {Vaiana} G.~S.,   {Harnden} F.~R.
  J.,  1992, \mn@doi [\apjs] {10.1086/191705}, \href
  {https://ui.adsabs.harvard.edu/#abs/1992ApJS...81..795G} {81, 795}

\bibitem[\protect\citeauthoryear{{Grimm}, {Gilfanov}  \& {Sunyaev}}{{Grimm}
  et~al.}{2002}]{Grimm02a}
{Grimm} H.-J.,  {Gilfanov} M.,   {Sunyaev} R.,  2002, \mn@doi [\aap]
  {10.1051/0004-6361:20020826}, \href
  {http://adsabs.harvard.edu/abs/2002A\%26A...391..923G} {391, 923}

\bibitem[\protect\citeauthoryear{{Grimm}, {Gilfanov}  \& {Sunyaev}}{{Grimm}
  et~al.}{2003}]{Grimm03a}
{Grimm} H.-J.,  {Gilfanov} M.,   {Sunyaev} R.,  2003, \mn@doi [\mnras]
  {10.1046/j.1365-8711.2003.06224.x}, \href
  {http://adsabs.harvard.edu/abs/2003MNRAS.339..793G} {339, 793}

\bibitem[\protect\citeauthoryear{{Grindlay}, {Petro}  \&
  {McClintock}}{{Grindlay} et~al.}{1984}]{1984ApJ...276..621G}
{Grindlay} J.~E.,  {Petro} L.~D.,   {McClintock} J.~E.,  1984, \mn@doi [\apj]
  {10.1086/161650}, \href
  {https://ui.adsabs.harvard.edu/#abs/1984ApJ...276..621G} {276, 621}

\bibitem[\protect\citeauthoryear{{Hakala}, {Ramsay}, {Muhli}, {Charles},
  {Hannikainen}, {Mukai}  \& {Vilhu}}{{Hakala}
  et~al.}{2005}]{2005MNRAS.356.1133H}
{Hakala} P.,  {Ramsay} G.,  {Muhli} P.,  {Charles} P.,  {Hannikainen} D.,
  {Mukai} K.,   {Vilhu} O.,  2005, \mn@doi [\mnras]
  {10.1111/j.1365-2966.2004.08543.x}, \href
  {https://ui.adsabs.harvard.edu/#abs/2005MNRAS.356.1133H} {356, 1133}

\bibitem[\protect\citeauthoryear{{Hjellming} \& {Johnston}}{{Hjellming} \&
  {Johnston}}{1981}]{1981ApJ...246L.141H}
{Hjellming} R.~M.,  {Johnston} K.~J.,  1981, \mn@doi [\apj] {10.1086/183571},
  \href {https://ui.adsabs.harvard.edu/#abs/1981ApJ...246L.141H} {246, L141}

\bibitem[\protect\citeauthoryear{{Hjellming} \& {Rupen}}{{Hjellming} \&
  {Rupen}}{1995}]{1995Natur.375..464H}
{Hjellming} R.~M.,  {Rupen} M.~P.,  1995, \mn@doi [\nat] {10.1038/375464a0},
  \href {http://adsabs.harvard.edu/abs/1995Natur.375..464H} {375, 464}

\bibitem[\protect\citeauthoryear{{Hogg}}{{Hogg}}{2018}]{Hogg18a}
{Hogg} D.~W.,  2018, preprint, \href
  {http://adsabs.harvard.edu/abs/2018arXiv180407766H} {} (\mn@eprint {arXiv}
  {1804.07766})

\bibitem[\protect\citeauthoryear{{Hutchings}, {Cowley}, {Crampton}, {van
  Paradijs}  \& {White}}{{Hutchings} et~al.}{1979}]{1979ApJ...229.1079H}
{Hutchings} J.~B.,  {Cowley} A.~P.,  {Crampton} D.,  {van Paradijs} J.,
  {White} N.~E.,  1979, \mn@doi [\apj] {10.1086/157042}, \href
  {https://ui.adsabs.harvard.edu/#abs/1979ApJ...229.1079H} {229, 1079}

\bibitem[\protect\citeauthoryear{{Ilovaisky}, {Chevalier}  \&
  {Motch}}{{Ilovaisky} et~al.}{1982}]{1982AA...114L...7I}
{Ilovaisky} S.~A.,  {Chevalier} C.,   {Motch} C.,  1982, \aap, \href
  {https://ui.adsabs.harvard.edu/#abs/1982A&A...114L...7I} {114, L7}

\bibitem[\protect\citeauthoryear{{Janka}}{{Janka}}{2013}]{Janka13a}
{Janka} H.-T.,  2013, \mn@doi [\mnras] {10.1093/mnras/stt1106}, \href
  {http://adsabs.harvard.edu/abs/2013MNRAS.434.1355J} {434, 1355}

\bibitem[\protect\citeauthoryear{{Janot-Pacheco}, {Ilovaisky}  \&
  {Chevalier}}{{Janot-Pacheco} et~al.}{1981}]{1981AA....99..274J}
{Janot-Pacheco} E.,  {Ilovaisky} S.~A.,   {Chevalier} C.,  1981, \aap, \href
  {https://ui.adsabs.harvard.edu/#abs/1981A&A....99..274J} {99, 274}

\bibitem[\protect\citeauthoryear{{Jonker} \& {Nelemans}}{{Jonker} \&
  {Nelemans}}{2004}]{Jonker04}
{Jonker} P.~G.,  {Nelemans} G.,  2004, \mn@doi [\mnras]
  {10.1111/j.1365-2966.2004.08193.x}, \href
  {http://adsabs.harvard.edu/abs/2004MNRAS.354..355J} {354, 355}

\bibitem[\protect\citeauthoryear{{Jonker}, {Galloway}, {McClintock}, {Buxton},
  {Garcia}  \& {Murray}}{{Jonker} et~al.}{2004}]{Jonker04a}
{Jonker} P.~G.,  {Galloway} D.~K.,  {McClintock} J.~E.,  {Buxton} M.,  {Garcia}
  M.,   {Murray} S.,  2004, \mn@doi [\mnras]
  {10.1111/j.1365-2966.2004.08246.x}, \href
  {http://adsabs.harvard.edu/abs/2004MNRAS.354..666J} {354, 666}

\bibitem[\protect\citeauthoryear{{Kaaret}, {Piraino}, {Halpern}  \&
  {Eracleous}}{{Kaaret} et~al.}{1999}]{1999ApJ...523..197K}
{Kaaret} P.,  {Piraino} S.,  {Halpern} J.,   {Eracleous} M.,  1999, \mn@doi
  [\apj] {10.1086/307711}, \href
  {https://ui.adsabs.harvard.edu/#abs/1999ApJ...523..197K} {523, 197}

\bibitem[\protect\citeauthoryear{{Kaluzienski}, {Holt}  \&
  {Swank}}{{Kaluzienski} et~al.}{1980}]{1980ApJ...241..779K}
{Kaluzienski} L.~J.,  {Holt} S.~S.,   {Swank} J.~H.,  1980, \mn@doi [\apj]
  {10.1086/158388}, \href
  {https://ui.adsabs.harvard.edu/#abs/1980ApJ...241..779K} {241, 779}

\bibitem[\protect\citeauthoryear{{Kawai} \& {Suzuki}}{{Kawai} \&
  {Suzuki}}{2005}]{2005ATel..534....1K}
{Kawai} N.,  {Suzuki} M.,  2005, The Astronomer's Telegram, \href
  {https://ui.adsabs.harvard.edu/#abs/2005ATel..534....1K} {534, 1}

\bibitem[\protect\citeauthoryear{{Koda} et~al.,}{{Koda} et~al.}{2012}]{Koda12a}
{Koda} J.,  et~al., 2012, \mn@doi [\apj] {10.1088/0004-637X/761/1/41}, \href
  {http://adsabs.harvard.edu/abs/2012ApJ...761...41K} {761, 41}

\bibitem[\protect\citeauthoryear{{Krimm} et~al.,}{{Krimm}
  et~al.}{2013}]{Krimm2013}
{Krimm} H.~A.,  et~al., 2013, \mn@doi [\apjs] {10.1088/0067-0049/209/1/14},
  209, 14

\bibitem[\protect\citeauthoryear{{Krivonos}, {Tsygankov}, {Revnivtsev},
  {Grebenev}, {Churazov}  \& {Sunyaev}}{{Krivonos} et~al.}{2010}]{Krivonos2010}
{Krivonos} R.,  {Tsygankov} S.,  {Revnivtsev} M.,  {Grebenev} S.,  {Churazov}
  E.,   {Sunyaev} R.,  2010, \mn@doi [\aap] {10.1051/0004-6361/201014935}, 523,
  A61

\bibitem[\protect\citeauthoryear{{Krzeminski}}{{Krzeminski}}{1974}]{1974ApJ...192L.135K}
{Krzeminski} W.,  1974, \mn@doi [\apj] {10.1086/181609}, \href
  {https://ui.adsabs.harvard.edu/#abs/1974ApJ...192L.135K} {192, L135}

\bibitem[\protect\citeauthoryear{{Kuntz}, {Long}  \& {Kilgard}}{{Kuntz}
  et~al.}{2016}]{Kuntz16}
{Kuntz} K.~D.,  {Long} K.~S.,   {Kilgard} R.~E.,  2016, \mn@doi [\apj]
  {10.3847/0004-637X/827/1/46}, \href
  {https://ui.adsabs.harvard.edu/#abs/2016ApJ...827...46K} {827, 46}

\bibitem[\protect\citeauthoryear{{Kuulkers}, {den Hartog}, {in't Zand},
  {Verbunt}, {Harris}  \& {Cocchi}}{{Kuulkers} et~al.}{2003}]{Kuulkers03a}
{Kuulkers} E.,  {den Hartog} P.~R.,  {in't Zand} J.~J.~M.,  {Verbunt} F.~W.~M.,
   {Harris} W.~E.,   {Cocchi} M.,  2003, \mn@doi [\aap]
  {10.1051/0004-6361:20021781}, \href
  {http://adsabs.harvard.edu/abs/2003A\%26A...399..663K} {399, 663}

\bibitem[\protect\citeauthoryear{{La Parola}, {Cusumano}, {Romano}, {Segreto},
  {Vercellone}  \& {Chincarini}}{{La Parola}
  et~al.}{2010}]{2010MNRAS.405L..66L}
{La Parola} V.,  {Cusumano} G.,  {Romano} P.,  {Segreto} A.,  {Vercellone} S.,
   {Chincarini} G.,  2010, \mn@doi [\mnras] {10.1111/j.1745-3933.2010.00860.x},
  \href {https://ui.adsabs.harvard.edu/abs/2010MNRAS.405L..66L} {405, L66}

\bibitem[\protect\citeauthoryear{{Leahy}}{{Leahy}}{2002}]{2002AA...391..219L}
{Leahy} D.~A.,  2002, \mn@doi [\aap] {10.1051/0004-6361:20020781}, \href
  {https://ui.adsabs.harvard.edu/abs/2002A&A...391..219L} {391, 219}

\bibitem[\protect\citeauthoryear{{Lehmer}, {Alexander}, {Bauer}, {Brandt},
  {Goulding}, {Jenkins}, {Ptak}  \& {Roberts}}{{Lehmer}
  et~al.}{2010}]{Lehmer10a}
{Lehmer} B.~D.,  {Alexander} D.~M.,  {Bauer} F.~E.,  {Brandt} W.~N.,
  {Goulding} A.~D.,  {Jenkins} L.~P.,  {Ptak} A.,   {Roberts} T.~P.,  2010,
  \mn@doi [\apj] {10.1088/0004-637X/724/1/559}, \href
  {http://adsabs.harvard.edu/abs/2010ApJ...724..559L} {724, 559}

\bibitem[\protect\citeauthoryear{{Lewin}, {van Paradijs}  \& {Taam}}{{Lewin}
  et~al.}{1993}]{Lewin93a}
{Lewin} W.~H.~G.,  {van Paradijs} J.,   {Taam} R.~E.,  1993, \mn@doi [\ssr]
  {10.1007/BF00196124}, \href
  {http://adsabs.harvard.edu/abs/1993SSRv...62..223L} {62, 223}

\bibitem[\protect\citeauthoryear{{Lin}, {Webb}  \& {Barret}}{{Lin}
  et~al.}{2012}]{Lin2012}
{Lin} D.,  {Webb} N.~A.,   {Barret} D.,  2012, \mn@doi [\apj]
  {10.1088/0004-637X/756/1/27}, 756, 27

\bibitem[\protect\citeauthoryear{{Linares}, {Shahbaz}  \& {Casares}}{{Linares}
  et~al.}{2018}]{Linares18a}
{Linares} M.,  {Shahbaz} T.,   {Casares} J.,  2018, \mn@doi [\apj]
  {10.3847/1538-4357/aabde6}, \href
  {http://adsabs.harvard.edu/abs/2018ApJ...859...54L} {859, 54}

\bibitem[\protect\citeauthoryear{{Lindegren} et~al.}{{Lindegren}
  et~al.}{2018}]{lindegren2018}
{Lindegren} L.,  et~al., 2018, \mn@doi [\aap] {10.1051/0004-6361/201832727},
  \href {https://ui.adsabs.harvard.edu/abs/2018A&A...616A...2L} {616, A2}

\bibitem[\protect\citeauthoryear{{Liu}, {van Paradijs}  \& {van den
  Heuvel}}{{Liu} et~al.}{2006}]{Liu06a}
{Liu} Q.~Z.,  {van Paradijs} J.,   {van den Heuvel} E.~P.~J.,  2006, \mn@doi
  [\aap] {10.1051/0004-6361:20064987}, \href
  {https://ui.adsabs.harvard.edu/#abs/2006A&A...455.1165L} {455, 1165}

\bibitem[\protect\citeauthoryear{{Liu}, {van Paradijs}  \& {van den
  Heuvel}}{{Liu} et~al.}{2007}]{Liu07a}
{Liu} Q.~Z.,  {van Paradijs} J.,   {van den Heuvel} E.~P.~J.,  2007, \mn@doi
  [\aap] {10.1051/0004-6361:20077303}, \href
  {https://ui.adsabs.harvard.edu/#abs/2007A&A...469..807L} {469, 807}

\bibitem[\protect\citeauthoryear{{Luri} et~al.,}{{Luri} et~al.}{2018}]{Luri18a}
{Luri} X.,  et~al., 2018, \mn@doi [\aap] {10.1051/0004-6361/201832964}, \href
  {http://adsabs.harvard.edu/abs/2018A\%26A...616A...9L} {616, A9}

\bibitem[\protect\citeauthoryear{{Lyuty} \& {Za{\u{i}}tseva}}{{Lyuty} \&
  {Za{\u{i}}tseva}}{2000}]{2000AstL...26....9L}
{Lyuty} V.~M.,  {Za{\u{i}}tseva} G.~V.,  2000, \mn@doi [Astronomy Letters]
  {10.1134/1.20364}, \href
  {https://ui.adsabs.harvard.edu/#abs/2000AstL...26....9L} {26, 9}

\bibitem[\protect\citeauthoryear{{MacDonald} et~al.,}{{MacDonald}
  et~al.}{2014}]{2014ApJ...784....2M}
{MacDonald} R. K.~D.,  et~al., 2014, \mn@doi [\apj]
  {10.1088/0004-637X/784/1/2}, \href
  {https://ui.adsabs.harvard.edu/abs/2014ApJ...784....2M} {784, 2}

\bibitem[\protect\citeauthoryear{{Marsden}, {Gruber}, {Heindl}, {Pelling}  \&
  {Rothschild}}{{Marsden} et~al.}{1998}]{1998ApJ...502L.129M}
{Marsden} D.,  {Gruber} D.~E.,  {Heindl} W.~A.,  {Pelling} M.~R.,
  {Rothschild} R.~E.,  1998, \mn@doi [\apj] {10.1086/311510}, \href
  {https://ui.adsabs.harvard.edu/#abs/1998ApJ...502L.129M} {502, L129}

\bibitem[\protect\citeauthoryear{{Masetti} et~al.,}{{Masetti}
  et~al.}{2002}]{2002AA...382..104M}
{Masetti} N.,  et~al., 2002, \mn@doi [\aap] {10.1051/0004-6361:20011543}, \href
  {https://ui.adsabs.harvard.edu/#abs/2002A&A...382..104M} {382, 104}

\bibitem[\protect\citeauthoryear{{Masetti} et~al.,}{{Masetti}
  et~al.}{2006a}]{2006AA...449.1139M}
{Masetti} N.,  et~al., 2006a, \mn@doi [\aap] {10.1051/0004-6361:20054332},
  \href {https://ui.adsabs.harvard.edu/abs/2006A&A...449.1139M} {449, 1139}

\bibitem[\protect\citeauthoryear{{Masetti}, {Orlandini}, {Palazzi}, {Amati}  \&
  {Frontera}}{{Masetti} et~al.}{2006b}]{2006AA...453..295M}
{Masetti} N.,  {Orlandini} M.,  {Palazzi} E.,  {Amati} L.,   {Frontera} F.,
  2006b, \mn@doi [\aap] {10.1051/0004-6361:20065025}, \href
  {https://ui.adsabs.harvard.edu/#abs/2006A&A...453..295M} {453, 295}

\bibitem[\protect\citeauthoryear{{Masetti} et~al.,}{{Masetti}
  et~al.}{2006c}]{2006AA...455...11M}
{Masetti} N.,  et~al., 2006c, \mn@doi [\aap] {10.1051/0004-6361:20065111},
  \href {https://ui.adsabs.harvard.edu/abs/2006A&A...455...11M} {455, 11}

\bibitem[\protect\citeauthoryear{{Masetti} et~al.,}{{Masetti}
  et~al.}{2009}]{2009AA...495..121M}
{Masetti} N.,  et~al., 2009, \mn@doi [\aap] {10.1051/0004-6361:200811322},
  \href {https://ui.adsabs.harvard.edu/abs/2009A&A...495..121M} {495, 121}

\bibitem[\protect\citeauthoryear{{Mason} \& {Cordova}}{{Mason} \&
  {Cordova}}{1982}]{1982ApJ...262..253M}
{Mason} K.~O.,  {Cordova} F.~A.,  1982, \mn@doi [\apj] {10.1086/160416}, \href
  {https://ui.adsabs.harvard.edu/#abs/1982ApJ...262..253M} {262, 253}

\bibitem[\protect\citeauthoryear{{Massey}, {Johnson}  \&
  {Degioia-Eastwood}}{{Massey} et~al.}{1995}]{1995ApJ...454..151M}
{Massey} P.,  {Johnson} K.~E.,   {Degioia-Eastwood} K.,  1995, \mn@doi [\apj]
  {10.1086/176474}, \href
  {https://ui.adsabs.harvard.edu/#abs/1995ApJ...454..151M} {454, 151}

\bibitem[\protect\citeauthoryear{{McBride} et~al.,}{{McBride}
  et~al.}{2006}]{2006AA...451..267M}
{McBride} V.~A.,  et~al., 2006, \mn@doi [\aap] {10.1051/0004-6361:20054239},
  \href {https://ui.adsabs.harvard.edu/abs/2006A&A...451..267M} {451, 267}

\bibitem[\protect\citeauthoryear{{McClintock}, {Remillard}  \&
  {Margon}}{{McClintock} et~al.}{1981}]{1981ApJ...243..900M}
{McClintock} J.~E.,  {Remillard} R.~A.,   {Margon} B.,  1981, \mn@doi [\apj]
  {10.1086/158655}, \href
  {https://ui.adsabs.harvard.edu/abs/1981ApJ...243..900M} {243, 900}

\bibitem[\protect\citeauthoryear{{Megier}, {Strobel}, {Galazutdinov}  \&
  {Kre{\l}owski}}{{Megier} et~al.}{2009}]{2009AA...507..833M}
{Megier} A.,  {Strobel} A.,  {Galazutdinov} G.~A.,   {Kre{\l}owski} J.,  2009,
  \mn@doi [\aap] {10.1051/0004-6361/20079144}, \href
  {https://ui.adsabs.harvard.edu/abs/2009A&A...507..833M} {507, 833}

\bibitem[\protect\citeauthoryear{{Merloni} et~al.,}{{Merloni}
  et~al.}{2012}]{Merloni12a}
{Merloni} A.,  et~al., 2012, preprint, \href
  {http://adsabs.harvard.edu/abs/2012arXiv1209.3114M} {} (\mn@eprint {arXiv}
  {1209.3114})

\bibitem[\protect\citeauthoryear{{Miller-Jones}, {Jonker}, {Dhawan}, {Brisken},
  {Rupen}, {Nelemans}  \& {Gallo}}{{Miller-Jones}
  et~al.}{2009}]{2009ApJ...706L.230M}
{Miller-Jones} J.~C.~A.,  {Jonker} P.~G.,  {Dhawan} V.,  {Brisken} W.,  {Rupen}
  M.~P.,  {Nelemans} G.,   {Gallo} E.,  2009, \mn@doi [\apj]
  {10.1088/0004-637X/706/2/L230}, \href
  {https://ui.adsabs.harvard.edu/#abs/2009ApJ...706L.230M} {706, L230}

\bibitem[\protect\citeauthoryear{{Mineo}, {Gilfanov}  \& {Sunyaev}}{{Mineo}
  et~al.}{2012}]{Mineo12a}
{Mineo} S.,  {Gilfanov} M.,   {Sunyaev} R.,  2012, \mn@doi [\mnras]
  {10.1111/j.1365-2966.2011.19862.x}, \href
  {http://adsabs.harvard.edu/abs/2012MNRAS.419.2095M} {419, 2095}

\bibitem[\protect\citeauthoryear{{Motch}, {Haberl}, {Dennerl}, {Pakull}  \&
  {Janot-Pacheco}}{{Motch} et~al.}{1997}]{1997AA...323..853M}
{Motch} C.,  {Haberl} F.,  {Dennerl} K.,  {Pakull} M.,   {Janot-Pacheco} E.,
  1997, \aap, \href {https://ui.adsabs.harvard.edu/#abs/1997A&A...323..853M}
  {323, 853}

\bibitem[\protect\citeauthoryear{{Mu{\~n}oz-Darias}, {Casares}  \&
  {Mart{\'{\i}}nez-Pais}}{{Mu{\~n}oz-Darias} et~al.}{2005}]{MunozDarias05a}
{Mu{\~n}oz-Darias} T.,  {Casares} J.,   {Mart{\'{\i}}nez-Pais} I.~G.,  2005,
  \mn@doi [\apj] {10.1086/497420}, \href
  {http://adsabs.harvard.edu/abs/2005ApJ...635..502M} {635, 502}

\bibitem[\protect\citeauthoryear{{Muno}, {Chakrabarty}, {Galloway}  \&
  {Savov}}{{Muno} et~al.}{2001}]{2001ApJ...553L.157M}
{Muno} M.~P.,  {Chakrabarty} D.,  {Galloway} D.~K.,   {Savov} P.,  2001,
  \mn@doi [\apj] {10.1086/320682}, \href
  {https://ui.adsabs.harvard.edu/#abs/2001ApJ...553L.157M} {553, L157}

\bibitem[\protect\citeauthoryear{{Muno}, {Pfahl}, {Baganoff}, {Brandt}, {Ghez},
  {Lu}  \& {Morris}}{{Muno} et~al.}{2005}]{Muno05a}
{Muno} M.~P.,  {Pfahl} E.,  {Baganoff} F.~K.,  {Brandt} W.~N.,  {Ghez} A.,
  {Lu} J.,   {Morris} M.~R.,  2005, \mn@doi [\apjl] {10.1086/429721}, \href
  {http://adsabs.harvard.edu/abs/2005ApJ...622L.113M} {622, L113}

\bibitem[\protect\citeauthoryear{{N{\"a}ttil{\"a}}, {Miller}, {Steiner},
  {Kajava}, {Suleimanov}  \& {Poutanen}}{{N{\"a}ttil{\"a}}
  et~al.}{2017}]{Nattila17a}
{N{\"a}ttil{\"a}} J.,  {Miller} M.~C.,  {Steiner} A.~W.,  {Kajava} J.~J.~E.,
  {Suleimanov} V.~F.,   {Poutanen} J.,  2017, \mn@doi [\aap]
  {10.1051/0004-6361/201731082}, \href
  {http://adsabs.harvard.edu/abs/2017A\%26A...608A..31N} {608, A31}

\bibitem[\protect\citeauthoryear{{Negueruela}, {Roche}, {Buckley},
  {Chakrabarty}, {Coe}, {Fabregat}  \& {Reig}}{{Negueruela}
  et~al.}{1996}]{1996AA...315..160N}
{Negueruela} I.,  {Roche} P.,  {Buckley} D.~A.~H.,  {Chakrabarty} D.,  {Coe}
  M.~J.,  {Fabregat} J.,   {Reig} P.,  1996, \aap, \href
  {https://ui.adsabs.harvard.edu/#abs/1996A&A...315..160N} {315, 160}

\bibitem[\protect\citeauthoryear{{Negueruela}, {Roche}, {Fabregat}  \&
  {Coe}}{{Negueruela} et~al.}{1999}]{1999MNRAS.307..695N}
{Negueruela} I.,  {Roche} P.,  {Fabregat} J.,   {Coe} M.~J.,  1999, \mn@doi
  [\mnras] {10.1046/j.1365-8711.1999.02682.x}, \href
  {https://ui.adsabs.harvard.edu/#abs/1999MNRAS.307..695N} {307, 695}

\bibitem[\protect\citeauthoryear{{Negueruela}, {Smith}, {Harrison}  \&
  {Torrej{\'o}n}}{{Negueruela} et~al.}{2006}]{2006ApJ...638..982N}
{Negueruela} I.,  {Smith} D.~M.,  {Harrison} T.~E.,   {Torrej{\'o}n} J.~M.,
  2006, \mn@doi [\apj] {10.1086/498935}, \href
  {https://ui.adsabs.harvard.edu/#abs/2006ApJ...638..982N} {638, 982}

\bibitem[\protect\citeauthoryear{{Parkes}, {Murdin}  \& {Mason}}{{Parkes}
  et~al.}{1980}]{1980MNRAS.190..537P}
{Parkes} G.~E.,  {Murdin} P.~G.,   {Mason} K.~O.,  1980, \mn@doi [\mnras]
  {10.1093/mnras/190.3.537}, \href
  {https://ui.adsabs.harvard.edu/#abs/1980MNRAS.190..537P} {190, 537}

\bibitem[\protect\citeauthoryear{{Pellizza}, {Chaty}  \&
  {Negueruela}}{{Pellizza} et~al.}{2006}]{2006AA...455..653P}
{Pellizza} L.~J.,  {Chaty} S.,   {Negueruela} I.,  2006, \mn@doi [\aap]
  {10.1051/0004-6361:20054436}, \href
  {https://ui.adsabs.harvard.edu/#abs/2006A&A...455..653P} {455, 653}

\bibitem[\protect\citeauthoryear{{Perryman} et~al.,}{{Perryman}
  et~al.}{1997}]{1997AA...323L..49P}
{Perryman} M.~A.~C.,  et~al., 1997, \aap, \href
  {https://ui.adsabs.harvard.edu/#abs/1997A&A...323L..49P} {323, L49}

\bibitem[\protect\citeauthoryear{{Pettitt}, {Ragan}  \& {Smith}}{{Pettitt}
  et~al.}{2020}]{Pettitt2020}
{Pettitt} A.~R.,  {Ragan} S.~E.,   {Smith} M.~C.,  2020, \mn@doi [\mnras]
  {10.1093/mnras/stz3155}, \href
  {https://ui.adsabs.harvard.edu/abs/2020MNRAS.491.2162P} {491, 2162}

\bibitem[\protect\citeauthoryear{{Phillips}, {Shahbaz}  \&
  {Podsiadlowski}}{{Phillips} et~al.}{1999}]{Phillips99a}
{Phillips} S.~N.,  {Shahbaz} T.,   {Podsiadlowski} P.,  1999, \mn@doi [\mnras]
  {10.1046/j.1365-8711.1999.02357.x}, \href
  {http://adsabs.harvard.edu/abs/1999MNRAS.304..839P} {304, 839}

\bibitem[\protect\citeauthoryear{{Podsiadlowski} \&
  {Rappaport}}{{Podsiadlowski} \& {Rappaport}}{2000}]{Podsiadlowski00a}
{Podsiadlowski} P.,  {Rappaport} S.,  2000, \mn@doi [\apj] {10.1086/308323},
  \href {http://adsabs.harvard.edu/abs/2000ApJ...529..946P} {529, 946}

\bibitem[\protect\citeauthoryear{{Polcaro} et~al.,}{{Polcaro}
  et~al.}{1990}]{1990AA...231..354P}
{Polcaro} V.~F.,  et~al., 1990, \aap, \href
  {https://ui.adsabs.harvard.edu/abs/1990A&A...231..354P} {231, 354}

\bibitem[\protect\citeauthoryear{{Pooley} et~al.,}{{Pooley}
  et~al.}{2003}]{Pooley03a}
{Pooley} D.,  et~al., 2003, \mn@doi [\apjl] {10.1086/377074}, \href
  {http://adsabs.harvard.edu/abs/2003ApJ...591L.131P} {591, L131}

\bibitem[\protect\citeauthoryear{{Pri{\v{s}}egen}}{{Pri{\v{s}}egen}}{2019}]{Prisegen2019}
{Pri{\v{s}}egen} M.,  2019, \mn@doi [\aap] {10.1051/0004-6361/201832682}, 621,
  A37

\bibitem[\protect\citeauthoryear{{Reid}, {McClintock}, {Narayan}, {Gou},
  {Remillard}  \& {Orosz}}{{Reid} et~al.}{2011a}]{2011ApJ...742...83R}
{Reid} M.~J.,  {McClintock} J.~E.,  {Narayan} R.,  {Gou} L.,  {Remillard}
  R.~A.,   {Orosz} J.~A.,  2011a, \mn@doi [\apj] {10.1088/0004-637X/742/2/83},
  \href {http://adsabs.harvard.edu/abs/2011ApJ...742...83R} {742, 83}

\bibitem[\protect\citeauthoryear{{Reid}, {McClintock}, {Narayan}, {Gou},
  {Remillard}  \& {Orosz}}{{Reid} et~al.}{2011b}]{Reid2011}
{Reid} M.~J.,  {McClintock} J.~E.,  {Narayan} R.,  {Gou} L.,  {Remillard}
  R.~A.,   {Orosz} J.~A.,  2011b, \mn@doi [\apj] {10.1088/0004-637X/742/2/83},
  742, 83

\bibitem[\protect\citeauthoryear{{Reig} \& {Fabregat}}{{Reig} \&
  {Fabregat}}{2015}]{Reig15a}
{Reig} P.,  {Fabregat} J.,  2015, \mn@doi [\aap] {10.1051/0004-6361/201425008},
  \href {http://adsabs.harvard.edu/abs/2015A\%26A...574A..33R} {574, A33}

\bibitem[\protect\citeauthoryear{{Reig}, {Chakrabarty}, {Coe}, {Fabregat},
  {Negueruela}, {Prince}, {Roche}  \& {Steele}}{{Reig}
  et~al.}{1996}]{1996AA...311..879R}
{Reig} P.,  {Chakrabarty} D.,  {Coe} M.~J.,  {Fabregat} J.,  {Negueruela} I.,
  {Prince} T.~A.,  {Roche} P.,   {Steele} I.~A.,  1996, \aap, \href
  {https://ui.adsabs.harvard.edu/#abs/1996A&A...311..879R} {311, 879}

\bibitem[\protect\citeauthoryear{{Reig}, {Negueruela}, {Papamastorakis},
  {Manousakis}  \& {Kougentakis}}{{Reig} et~al.}{2005}]{2005AA...440..637R}
{Reig} P.,  {Negueruela} I.,  {Papamastorakis} G.,  {Manousakis} A.,
  {Kougentakis} T.,  2005, \mn@doi [\aap] {10.1051/0004-6361:20052684}, \href
  {https://ui.adsabs.harvard.edu/#abs/2005A&A...440..637R} {440, 637}

\bibitem[\protect\citeauthoryear{{Reig}, {Zezas}  \& {Gkouvelis}}{{Reig}
  et~al.}{2010}]{2010AA...522A.107R}
{Reig} P.,  {Zezas} A.,   {Gkouvelis} L.,  2010, \mn@doi [\aap]
  {10.1051/0004-6361/201014788}, \href
  {https://ui.adsabs.harvard.edu/abs/2010A&A...522A.107R} {522, A107}

\bibitem[\protect\citeauthoryear{{Repetto}, {Davies}  \&
  {Sigurdsson}}{{Repetto} et~al.}{2012}]{Repetto12a}
{Repetto} S.,  {Davies} M.~B.,   {Sigurdsson} S.,  2012, \mn@doi [\mnras]
  {10.1111/j.1365-2966.2012.21549.x}, \href
  {http://adsabs.harvard.edu/abs/2012MNRAS.425.2799R} {425, 2799}

\bibitem[\protect\citeauthoryear{{Reynolds}, {Bell}  \& {Hilditch}}{{Reynolds}
  et~al.}{1992}]{1992MNRAS.256..631R}
{Reynolds} A.~P.,  {Bell} S.~A.,   {Hilditch} R.~W.,  1992, \mn@doi [\mnras]
  {10.1093/mnras/256.3.631}, \href
  {https://ui.adsabs.harvard.edu/#abs/1992MNRAS.256..631R} {256, 631}

\bibitem[\protect\citeauthoryear{{Reynolds}, {Quaintrell}, {Still}, {Roche},
  {Chakrabarty}  \& {Levine}}{{Reynolds} et~al.}{1997}]{1997MNRAS.288...43R}
{Reynolds} A.~P.,  {Quaintrell} H.,  {Still} M.~D.,  {Roche} P.,  {Chakrabarty}
  D.,   {Levine} S.~E.,  1997, \mn@doi [\mnras] {10.1093/mnras/288.1.43}, \href
  {https://ui.adsabs.harvard.edu/#abs/1997MNRAS.288...43R} {288, 43}

\bibitem[\protect\citeauthoryear{{Sadakane}, {Hirata}, {Jugaku}, {Kondo},
  {Matsuoka}, {Tanaka}  \& {Hammerschlag-Hensberge}}{{Sadakane}
  et~al.}{1985}]{1985ApJ...288..284S}
{Sadakane} K.,  {Hirata} R.,  {Jugaku} J.,  {Kondo} Y.,  {Matsuoka} M.,
  {Tanaka} Y.,   {Hammerschlag-Hensberge} G.,  1985, \mn@doi [\apj]
  {10.1086/162791}, \href
  {https://ui.adsabs.harvard.edu/#abs/1985ApJ...288..284S} {288, 284}

\bibitem[\protect\citeauthoryear{{Samus'}, {Kazarovets}, {Durlevich}, {Kireeva}
   \& {Pastukhova}}{{Samus'} et~al.}{2017}]{Samus2017}
{Samus'} N.~N.,  {Kazarovets} E.~V.,  {Durlevich} O.~V.,  {Kireeva} N.~N.,
  {Pastukhova} E.~N.,  2017, \mn@doi [Astronomy Reports]
  {10.1134/S1063772917010085}, 61, 80

\bibitem[\protect\citeauthoryear{{Smith}}{{Smith}}{2004}]{2004ATel..338....1S}
{Smith} D.~M.,  2004, The Astronomer's Telegram, \href
  {https://ui.adsabs.harvard.edu/#abs/2004ATel..338....1S} {338, 1}

\bibitem[\protect\citeauthoryear{{Steele}, {Negueruela}, {Coe}  \&
  {Roche}}{{Steele} et~al.}{1998}]{1998MNRAS.297L...5S}
{Steele} I.~A.,  {Negueruela} I.,  {Coe} M.~J.,   {Roche} P.,  1998, \mn@doi
  [\mnras] {10.1046/j.1365-8711.1998.01593.x}, \href
  {https://ui.adsabs.harvard.edu/#abs/1998MNRAS.297L...5S} {297, L5}

\bibitem[\protect\citeauthoryear{{Steiner}, {Heinke}, {Bogdanov}, {Li}, {Ho},
  {Bahramian}  \& {Han}}{{Steiner} et~al.}{2018}]{Steiner18a}
{Steiner} A.~W.,  {Heinke} C.~O.,  {Bogdanov} S.,  {Li} C.~K.,  {Ho} W.~C.~G.,
  {Bahramian} A.,   {Han} S.,  2018, \mn@doi [\mnras] {10.1093/mnras/sty215},
  \href {http://adsabs.harvard.edu/abs/2018MNRAS.476..421S} {476, 421}

\bibitem[\protect\citeauthoryear{{Stevens}, {Reig}, {Coe}, {Buckley},
  {Fabregat}  \& {Steele}}{{Stevens} et~al.}{1997}]{1997MNRAS.288..988S}
{Stevens} J.~B.,  {Reig} P.,  {Coe} M.~J.,  {Buckley} D.~A.~H.,  {Fabregat} J.,
    {Steele} I.~A.,  1997, \mn@doi [\mnras] {10.1093/mnras/288.4.988}, \href
  {https://ui.adsabs.harvard.edu/abs/1997MNRAS.288..988S} {288, 988}

\bibitem[\protect\citeauthoryear{{Strohmayer} \& {Bildsten}}{{Strohmayer} \&
  {Bildsten}}{2006}]{Strohmayer06a}
{Strohmayer} T.,  {Bildsten} L.,  2006, in {Lewin} W.~H.~G.,  {van der Klis}
  M.,  eds, Compact stellar X-ray sources. Cambridge Astrophysics Series.
Cambridge University Press, pp 113--156, \mn@doi{10.2277/0521826594}

\bibitem[\protect\citeauthoryear{{Swartz}, {Ghosh}, {McCollough}, {Pannuti},
  {Tennant}  \& {Wu}}{{Swartz} et~al.}{2003}]{Swartz03a}
{Swartz} D.~A.,  {Ghosh} K.~K.,  {McCollough} M.~L.,  {Pannuti} T.~G.,
  {Tennant} A.~F.,   {Wu} K.,  2003, \mn@doi [\apjs] {10.1086/345084}, \href
  {http://adsabs.harvard.edu/abs/2003ApJS..144..213S} {144, 213}

\bibitem[\protect\citeauthoryear{{Tetarenko}, {Sivakoff}, {Heinke}  \&
  {Gladstone}}{{Tetarenko} et~al.}{2016}]{Tetarenko16a}
{Tetarenko} B.~E.,  {Sivakoff} G.~R.,  {Heinke} C.~O.,   {Gladstone} J.~C.,
  2016, \mn@doi [\apjs] {10.3847/0067-0049/222/2/15}, \href
  {http://adsabs.harvard.edu/abs/2016ApJS..222...15T} {222, 15}

\bibitem[\protect\citeauthoryear{{Th{\'e}venin}, {Falanga}, {Kuo},
  {Pietrzy{\'n}ski}  \& {Yamaguchi}}{{Th{\'e}venin}
  et~al.}{2017}]{Thevenin2017}
{Th{\'e}venin} F.,  {Falanga} M.,  {Kuo} C.~Y.,  {Pietrzy{\'n}ski} G.,
  {Yamaguchi} M.,  2017, \mn@doi [\ssr] {10.1007/s11214-017-0418-9}, \href
  {https://ui.adsabs.harvard.edu/abs/2017SSRv..212.1787T} {212, 1787}

\bibitem[\protect\citeauthoryear{{Tomsick}, {Gelino}  \& {Kaaret}}{{Tomsick}
  et~al.}{2005}]{2005ApJ...635.1233T}
{Tomsick} J.~A.,  {Gelino} D.~M.,   {Kaaret} P.,  2005, \mn@doi [\apj]
  {10.1086/497587}, \href
  {https://ui.adsabs.harvard.edu/#abs/2005ApJ...635.1233T} {635, 1233}

\bibitem[\protect\citeauthoryear{{Torrej{\'o}n} \& {Orr}}{{Torrej{\'o}n} \&
  {Orr}}{2001}]{Torrejon01a}
{Torrej{\'o}n} J.~M.,  {Orr} A.,  2001, \mn@doi [\aap]
  {10.1051/0004-6361:20011070}, \href
  {https://ui.adsabs.harvard.edu/#abs/2001A&A...377..148T} {377, 148}

\bibitem[\protect\citeauthoryear{{Torrej{\'o}n}, {Negueruela}, {Smith}  \&
  {Harrison}}{{Torrej{\'o}n} et~al.}{2010}]{2010AA...510A..61T}
{Torrej{\'o}n} J.~M.,  {Negueruela} I.,  {Smith} D.~M.,   {Harrison} T.~E.,
  2010, \mn@doi [\aap] {10.1051/0004-6361/200912619}, \href
  {https://ui.adsabs.harvard.edu/abs/2010A&A...510A..61T} {510, A61}

\bibitem[\protect\citeauthoryear{{Tremmel} et~al.,}{{Tremmel}
  et~al.}{2013}]{Tremmel13a}
{Tremmel} M.,  et~al., 2013, \mn@doi [\apj] {10.1088/0004-637X/766/1/19}, \href
  {http://adsabs.harvard.edu/abs/2013ApJ...766...19T} {766, 19}

\bibitem[\protect\citeauthoryear{{Tsygankov} \& {Lutovinov}}{{Tsygankov} \&
  {Lutovinov}}{2005}]{2005AstL...31...88T}
{Tsygankov} S.~S.,  {Lutovinov} A.~A.,  2005, \mn@doi [Astronomy Letters]
  {10.1134/1.1862348}, \href
  {https://ui.adsabs.harvard.edu/abs/2005AstL...31...88T} {31, 88}

\bibitem[\protect\citeauthoryear{{Vall{\'e}e}}{{Vall{\'e}e}}{2008}]{Vallee08a}
{Vall{\'e}e} J.~P.,  2008, \mn@doi [\aj] {10.1088/0004-6256/135/4/1301}, \href
  {http://adsabs.harvard.edu/abs/2008AJ....135.1301V} {135, 1301}

\bibitem[\protect\citeauthoryear{{Vall{\'e}e}}{{Vall{\'e}e}}{2014}]{Vallee14a}
{Vall{\'e}e} J.~P.,  2014, \aj, 148, 5

\bibitem[\protect\citeauthoryear{{Vall{\'e}e}}{{Vall{\'e}e}}{2017}]{Vallee17a}
{Vall{\'e}e} J.~P.,  2017, \mn@doi [The Astronomical Review]
  {10.1080/21672857.2017.1379459}, \href
  {http://adsabs.harvard.edu/abs/2017AstRv..13..113V} {13, 113}

\bibitem[\protect\citeauthoryear{{Verbunt}}{{Verbunt}}{2003}]{Verbunt03a}
{Verbunt} F.,  2003, in {Piotto} G.,  {Meylan} G.,  {Djorgovski} S.~G.,
  {Riello} M.,  eds,  Astronomical Society of the Pacific Conference Series
  Vol. 296, New Horizons in Globular Cluster Astronomy. p.~245 (\mn@eprint {}
  {astro-ph/0210057})

\bibitem[\protect\citeauthoryear{{Verbunt} \& {Hut}}{{Verbunt} \&
  {Hut}}{1987}]{Verbunt87a}
{Verbunt} F.,  {Hut} P.,  1987, in {Helfand} D.~J.,  {Huang} J.-H.,  eds,  IAU
  Symposium Vol. 125, The Origin and Evolution of Neutron Stars. p.~187

\bibitem[\protect\citeauthoryear{{Verbunt} \& {Lewin}}{{Verbunt} \&
  {Lewin}}{2006}]{Verbunt06a}
{Verbunt} F.,  {Lewin} W.~H.~G.,  2006, in {Lewin} W.~H.~G.,  {van der Klis}
  M.,  eds, Compact stellar X-ray sources. Cambridge Astrophysics Series.
Cambridge University Press, pp 341--379, \mn@doi{10.2277/0521826594}

\bibitem[\protect\citeauthoryear{{Verbunt}, {van Paradijs}  \&
  {Elson}}{{Verbunt} et~al.}{1984}]{Verbunt84a}
{Verbunt} F.,  {van Paradijs} J.,   {Elson} R.,  1984, \mn@doi [\mnras]
  {10.1093/mnras/210.4.899}, \href
  {http://cdsads.u-strasbg.fr/abs/1984MNRAS.210..899V} {210, 899}

\bibitem[\protect\citeauthoryear{{Wachter} \& {Smale}}{{Wachter} \&
  {Smale}}{1998}]{1998ApJ...496L..21W}
{Wachter} S.,  {Smale} A.~P.,  1998, \mn@doi [\apj] {10.1086/311242}, \href
  {https://ui.adsabs.harvard.edu/#abs/1998ApJ...496L..21W} {496, L21}

\bibitem[\protect\citeauthoryear{{Wen}, {Remillard}  \& {Bradt}}{{Wen}
  et~al.}{2000}]{2000ApJ...532.1119W}
{Wen} L.,  {Remillard} R.~A.,   {Bradt} H.~V.,  2000, \mn@doi [\apj]
  {10.1086/308604}, \href
  {https://ui.adsabs.harvard.edu/#abs/2000ApJ...532.1119W} {532, 1119}

\bibitem[\protect\citeauthoryear{Wenger et~al.,}{Wenger
  et~al.}{2000}]{wenger_simbad_2000}
Wenger M.,  et~al., 2000, \mn@doi [\aaps] {10.1051/aas:2000332}, 143, 9

\bibitem[\protect\citeauthoryear{{Wilson}, {Finger}, {Coe}, {Laycock}  \&
  {Fabregat}}{{Wilson} et~al.}{2002}]{2002ApJ...570..287W}
{Wilson} C.~A.,  {Finger} M.~H.,  {Coe} M.~J.,  {Laycock} S.,   {Fabregat} J.,
  2002, \mn@doi [\apj] {10.1086/339739}, \href
  {https://ui.adsabs.harvard.edu/#abs/2002ApJ...570..287W} {570, 287}

\bibitem[\protect\citeauthoryear{{Wilson}, {Finger}, {Coe}  \&
  {Negueruela}}{{Wilson} et~al.}{2003}]{2003ApJ...584..996W}
{Wilson} C.~A.,  {Finger} M.~H.,  {Coe} M.~J.,   {Negueruela} I.,  2003,
  \mn@doi [\apj] {10.1086/345791}, \href
  {https://ui.adsabs.harvard.edu/#abs/2003ApJ...584..996W} {584, 996}

\bibitem[\protect\citeauthoryear{{Zhang}, {Gilfanov}  \& {Bogd{\'a}n}}{{Zhang}
  et~al.}{2012}]{Zhang12a}
{Zhang} Z.,  {Gilfanov} M.,   {Bogd{\'a}n} {\'A}.,  2012, \mn@doi [\aap]
  {10.1051/0004-6361/201219015}, \href
  {http://adsabs.harvard.edu/abs/2012A\%26A...546A..36Z} {546, A36}

\bibitem[\protect\citeauthoryear{{in't Zand} et~al.,}{{in't Zand}
  et~al.}{2002}]{2002AA...389L..43I}
{in't Zand} J.~J.~M.,  et~al., 2002, \mn@doi [\aap]
  {10.1051/0004-6361:20020631}, \href
  {https://ui.adsabs.harvard.edu/#abs/2002A&A...389L..43I} {389, L43}

\bibitem[\protect\citeauthoryear{{in't Zand}, {Cumming}, {van der Sluys},
  {Verbunt}  \& {Pols}}{{in't Zand} et~al.}{2005}]{2005AA...441..675I}
{in't Zand} J.~J.~M.,  {Cumming} A.,  {van der Sluys} M.~V.,  {Verbunt} F.,
  {Pols} O.~R.,  2005, \mn@doi [\aap] {10.1051/0004-6361:20053002}, \href
  {https://ui.adsabs.harvard.edu/#abs/2005A&A...441..675I} {441, 675}

\bibitem[\protect\citeauthoryear{{van Paradijs}}{{van
  Paradijs}}{1978}]{vanParadijs78a}
{van Paradijs} J.,  1978, \mn@doi [\nat] {10.1038/274650a0}, \href
  {http://cdsads.u-strasbg.fr/abs/1978Natur.274..650V} {274, 650}

\bibitem[\protect\citeauthoryear{{van Paradijs}}{{van
  Paradijs}}{1981}]{vanParadijs81a}
{van Paradijs} J.,  1981, \aap, \href
  {http://cdsads.u-strasbg.fr/abs/1981A\%26A...101..174V} {101, 174}

\bibitem[\protect\citeauthoryear{{van Paradijs}}{{van
  Paradijs}}{1998}]{vanParadijs98a}
{van Paradijs} J.,  1998, in {Buccheri} R.,  {van Paradijs} J.,   {Alpar} A.,
  eds,  NATO Advanced Science Institutes (ASI) Series C Vol. 515, NATO Advanced
  Science Institutes (ASI) Series C. p.~279 (\mn@eprint {} {astro-ph/9802177})

\bibitem[\protect\citeauthoryear{{van Paradijs} \& {White}}{{van Paradijs} \&
  {White}}{1995}]{1995ApJ...447L..33V}
{van Paradijs} J.,  {White} N.,  1995, \mn@doi [\apjl] {10.1086/309558}, \href
  {https://ui.adsabs.harvard.edu/abs/1995ApJ...447L..33V} {447, L33}

\makeatother
\end{thebibliography}

\appendix

\section{Updated distances and classifications of XRBs with \gaia counterparts}
\label{sect:apdxA}

For the 88 \gaia candidate counterparts to the Liu XRB sample (see \S\ref{sect:xmatch}), we searched the literature for more recently-published distances and 
compilations \citep{Tetarenko16a,blackcat2016, wenger_simbad_2000} for updates to classifications.

Two Liu catalogue objects have controversial classifications but do not figure in our analysis because their \gaia DR2 counterparts have no parallax.
SIMBAD notes that the nature of 2S~0053+604 ($\gamma$ Cas) as an X-ray binary is controversial \citep[for a summary, see][]{Prisegen2019}.
Although the star itself is in \gaia DR2, its bright magnitude ($G=1.82$) means that its observations require special processing expected in a later data release \citep{gaiaDR2}.
The object designated by \citet{Liu06a} as Swift J061223.0+701243.9 is claimed by SIMBAD to have incorrect nomenclature.
As far as we can tell, this object is real and correctly designated by Liu but it is not particularly well-studied, with no published distance estimate.
The most recent analysis is by \citet{Butters2011} who conclude that Swift J061223.0+701243.9  is probably an intermediate polar, but an X-ray binary nature cannot be ruled out.

Six objects listed by \citet{Liu06a} as LMXBs are classified by SIMBAD as HMXBs.
For three of these (1A~0620-00, GS~1124-684, GS~2023+338) the reference for the HMXB classification is \citet{Tetarenko16a}; however that catalogue does not give explicit LMXB/HMXB classifications.
SIMBAD lists 3A~1516-569 (Cir~X-1) and 3A~1954+319 as being classified as LMXBs by \citet{Baumgartner2013} and as HMXBs
by \citet{Samus2017} and \citet{Krivonos2010} respectively.
Neither of the latter two sources gives a reference or justification for the HMXB classification.
SIMBAD lists GRO~J1655-40  as being classified as HMXB by \citet{Lin2012} and LMXB by \citet{Krimm2013}.
However, \citet{Lin2012} did not classify sources as high- or low-mass XRBs, and the classifications in \citet{Krimm2013} are cited as originating from the literature or SIMBAD itself.
With no strong reasons to reclassify these six objects, we retain them in our list of LMXBs.

We were able to find published distance estimates for ten objects that
had no distance estimates listed by \citet{Liu06a,Liu07a}.
Fifteen additional objects in our sample had distance determinations more recent than those listed by \citet{Liu06a,Liu07a}.
We tabulate these in \autoref{tab:newdist} and use them in our analysis in \autoref{sec:dist_comp}.

Two objects in our sample have controversial distances: Cyg~X-1 and GRO~J1655-40.
The discrepancy between radio parallax distance \citep[from][]{Reid2011} and optical parallax from \gaia of Cyg X-1 is peculiar, as this system is one of the closest and brightest X-ray binaries (both in radio and optical). This apparent tension is likely caused by impact of the radio jet on the radio parallax (Miller-Jones et al., in prep).
The \gaia distance is more consistent with that reported in the Liu catalogue \citep[2.14~kpc;][]{1995ApJ...454..151M}.
\citet{2006A&A...457..249F} challenged the accepted distance to GRO~J1655-40 of 3.2~kpc, finding a distance of 1.7~kpc. 
Despite their strong claim, these authors show in their table 1 that the uncertainty in spectral class allows the upper limit on distance to be as high as 3--4 kpc (e.g., if the companion is F7\ion{ii}).
Interestingly, the \gaia counterpart parallax is consistent with the larger distance. 
However, the location of the source makes distance calculation based on parallax strongly dependent on the prior model:
there is a large discrepancy between the distance based on Bailer-Jones prior ($\sim3$~kpc) and one that considers distribution of BHs in the Milky Way ($\sim6-7$ kpc), as shown by \citet{Atri19}.

\begin{table}
\caption{
XRBs with \gaia candidate counterparts and newer
published distances
\label{tab:newdist}
} 
\begin{tabular}{lllll}
\hline
Name & Liu dist  & type & d$_{ \mathrm{prev}}$ (newer)  & type\\
& kpc & & kpc & \\
\hline
\hline
LMXBs & & & &\\
GRO J0422+32&	\nodata&	\nodata	&2.49&	SEDfit\\
1A 0620-00	&1.16&	phot&1.06	&phot\\
GS 1124-684	&5.5&	phot&5.9	&SEDfit\\
3A 1516-569	&\nodata	&\nodata	&9.2&	burst\\
GRO J1655-40&	1.7&	phot&	3.2&	jetPM\\
3A 1702-363	&\nodata	&\nodata&	9.2&unclear\\
4U 1724-307	&7	&burst&	9.5	&cluster\\
SLX 1737-282&	7.5&	burst	&6.5&	burst\\
4U 1908+005	&5	&Roche	&5.2	&burst\\
HMXBs & & & &\\
IGR J01583+6713	&\nodata	&\nodata	&6.4&	phot\\
EXO 051910+3737.7 &\nodata &\nodata&	1.7 &	phot\\
IGR J06074+2205	&\nodata	&\nodata	&4.5	&phot\\
XTE J0658-073	&\nodata&\nodata& 3.9	& phot	\\	
IGR J11215-5952	&8	&phot	&6.2&	phot\\
IGR J11435-6109	&\nodata&\nodata& 8.6	& phot	\\	
2S 1145-619	&2.25	&phot	&3.1	&phot\\
4U 1223-624	&5&	phot	&4.1	&phot\\
1H 1249-637	&0.3&	HipPLX	&0.392&	Av\\
IGR J16479-4514	&\nodata&\nodata& 4.45	& phot \\
4U 1700-37	&1.9&	phot	&2.12&	Av\\
IGR J17544-2619&	10	&unknown&	3.2&	phot\\
SAX J1819.3-2525&	6	&SEDfit	&6.2	&SEDfit\\
IGR J19140+0951	&\nodata&\nodata& 3.6	& phot	\\
KS 1947+300	&10&	phot	&9.5&	pulsar\\
4U 1956+35	&2.14&	cluster	&1.86&	VLBAPLX\\
\hline
\end{tabular}
{'Newer' here means non-\gaia DR2 distances published after the Liu et al. catalogues.}
\end{table}

\section{Distance discrepancies}
\label{sect:apdxB}

In this section we discuss five objects with large discrepancies between distances gathered from the literature and measured from \gaia DR2 with the
\citet{BailerJones18a} prior. Here we define `large' as 
$ |d_{ \mathrm{Gaia}}-d_{ \mathrm{prev}}|/(0.5 \times (d_{ \mathrm{Gaia}} + d_{ \mathrm{prev}})) > 1$. 
For all of these objects, the previously-published distance is well outside the \gaia low-to-high range.
We report the \gaia DR2 {\tt astrometric\_gof\_al} value as GOF. 
This quantity is expected to follow a normal distribution with zero mean and unit standard deviation; hence absolute values $\gtrsim 3$ indicate a poor fit.

\textbf{4U 2129+47/V1727 Cyg}: \gaia distance 1.75~kpc, GOF 1.92. The \citet{Liu07a} distance for this object is from the work of \citet{1990AJ.....99..678C}, who derive a distance of 6.3~kpc to the optical companion. Those authors mention that it is unclear that the companion and XRB are a true physical association, and that previous distance estimates to the XRB system generally give smaller distances \citep[e.g., 2.2 kpc; ][]{1981ApJ...243..900M}. We conclude that the \gaia distance is consistent with these earlier estimates.

\textbf{IGR J16318-4848}: \gaia distance 5.22~kpc, GOF 32.4. The \citet{Liu06a} distance for this object is from the work of \citet{2004ApJ...616..469F} who give a range of distances between 0.9 and 6.2 kpc, derived from SED fitting. A more recent work \citep{2020ApJ...894...86F} determines a distance from \gaia matching and derives the same distance as our work.
We conclude that the \gaia distance, although imprecise, is consistent with the broad range in the previous estimate.

\textbf{IGR J16465-4507}: \gaia distance 2.70~kpc, GOF 9.0. The \citet{Liu06a} distance for this object is from the work of \citet{2004ATel..338....1S} who give an estimated distance of 12.5 kpc based on photometry of the companion. The discussion of this object by \citet{2010MNRAS.405L..66L} explains that the optical companion is highly absorbed; Optical spectroscopic studies also provide additional evidence reaffirming the optical counterpart. The tension between the \gaia and previous distance estimates remains unresolved.

\textbf{XTE J1906+09}: \gaia distance 2.77~kpc, GOF 6.8. The \citet{Liu06a} distance distance for this object is from the work of \citet{1998ApJ...502L.129M} who give an estimate distance of 10 kpc based on neutral hydrogen absorption. 
However, 3D dust maps in this directions \citep{2019ApJ...887...93G} indicate that $E(g-r)\leq2.2$, which would suggest that the Galactic hydrogen column density in this direction is $\leq 2\times10^{22}$ cm$^{-2}$ \citep{Bahramian2015, Foight2016}. Thus we conclude that the \gaia distance is likely more reliable for this object. 

\textbf{KS 1947+300}: \gaia distance 3.1 kpc, GOF 0.0. The \citet{Liu06a} distance distance for this object is from the work of \citet{2005AstL...31...88T} who give an estimate distance of 9.5 kpc based on its X-ray pulsation properties.
While the \gaia fit appears good, it is important to note that the measured parallax is insignificant when uncertainties are considered.

\bsp	
\label{lastpage}
\end{document}